\documentclass[10pt]{article}

\usepackage{epsfig}
\usepackage{amssymb}
\usepackage{amsmath}
\usepackage{amsthm}
\usepackage{verbatim}
\usepackage{color}
\usepackage{makecell}
\newtheorem{definition}{Definition}
\usepackage{makecell}

\usepackage{subcaption}
\usepackage{multirow}
\usepackage{comment}

\renewcommand{\baselinestretch}{2.0}

\begin{document}

\pagestyle{plain} \pagenumbering{arabic} \setcounter{page}{1}
\vspace*{-1.5cm}\hspace*{9.0cm} 

\begin{center}

{\Large Historical Credibility for Movie Reviews and Its Application to Weakly Supervised Classification}

Min-Seon Kim\textsuperscript{1},
Bo-Young Lim\textsuperscript{1},
Han-Sub Shin\textsuperscript{2},
Hyuk-Yoon Kwon\textsuperscript{3} \text{*}
\\
\bigskip
\textbf{1} Department of Industrial Engineering, Seoul National University of Science and Technology, 232 Gongneung-Ro, Nowon-Gu, Seoul 01811, Korea
\\
\textbf{2} Graduate School of Data Science, Seoul National University of Science and Technology, 232 Gongneung-Ro, Nowon-Gu, Seoul 01811, Korea
\\
\textbf{3} Department of Industrial Engineering and Graduate School of Data Science, Seoul National University of Science and Technology, 232 Gongneung-Ro, Nowon-Gu, Seoul 01811, Korea

Corresponding author: hyukyoon.kwon@seoultech.ac.kr$^{*}$

\end{center}


\begin{abstract}
In this study, we deal with the problem of judging the credibility of movie reviews. The problem is challenging because even experts cannot clearly and efficiently judge the credibility of a movie review and the number of movie reviews is very large. To tackle this problem, we propose \textit{historical credibility} that judges the credibility of reviews based on the historical  ratings and textual reviews written by each reviewer. For this, we present three kinds of criteria that can clearly classify the reviews into trusted or distrusted ones. We validate the effectiveness of the proposed historical credibility through extensive analysis. Specifically, we show that characteristics between the trusted or distrusted reviews are quite distinguishable in terms of three viewpoints: 1) distribution, 2) statistics, and 3) correlation. Then, we apply historical credibility to a weakly supervised model to classify a given review as a trusted or distrusted one. First, we show that it is significantly efficient because the entire data set is annotated according to the predefined criteria. Indeed, it can annotate 6,400 movie reviews only in 0.093 seconds, which occupy only 0.55\%$\sim$1.88\% of the total learning time when we use LSTM and SVM as the learning model. Second, we show that the historical credibility-based classification model clearly outperforms the textual review-based classification model. Specifically, the classification accuracy of the former outperforms that of the latter by up to 11.7\%$\sim$13.4\%. In addition, we clearly confirm that our classification model shows higher accuracy as the data size increases.


\end{abstract}

\renewcommand{\baselinestretch}{2.0}

\vspace*{-0.5cm}




\section{Introduction}
\label{sec:sec1}

A movie review is a subjective evaluation written by a moviegoer, usually consisting of two kinds of data: 1) textual reviews and 2) ratings. Textual reviews are qualitative evaluations; ratings are quantitative evaluations expressed by scores. Table~\ref{tab:tab17} shows a sample movie review from IMDb (https://www.imdb.com/), a leading website managing movie reviews. The example shows the rating and the textual review for the movie. We note that these two types of information are closely related: the rating can be considered a numerical summarization reflecting the textual review.

\begin{table}[!h]
\fontsize{9}{10}\selectfont
\center{
\caption{Sample movie review} 
\begin{tabular}{|c|c|c|c|c|}

\hline
\textbf{Textual reviews} & \textbf{Rating}  & \textbf{User} & \textbf{Written Date} & \textbf{\makecell{Helpfulness votes / \\ Total votes}} \\
\hline
\hline \makecell{I enjoyed every second \\ of it. I even went to \\ the theater twice and \\came out with a big smile \\ on my face each time.} & 10 & lott**** & 16 July, 2019 & 19 / 26 \\

\hline
\end{tabular}
\label{tab:tab17}
}
\end{table}

Movie reviews have a significant impact on the decision-making of potential moviegoers~\cite{chintagunta2010effects}. However, a certain group of movie reviewers may generate indistinguishable or intentionally malicious reviews, biasing the overall movie rating. Although such groups ought to lack credibility, many well-known existing movie review management systems including Metacritic~\footnote{https://www.metacritic.com/} and Rotten Tomatoes~\footnote{https://www.rottentomatoes.com} do not consider the credibility of movie reviewers in calculating each movie's overall rating.

To enhance the effective evaluation of movies, it is necessary to consider the credibility of movie reviews in calculating the movie's rating. For this, we need a method to judge movie reviews according to their credibility. However, this is a challenging issue because even experts cannot clearly judge the movie review's credibility and the number of movie reviews is very large, which requires much time and effort for manual annotation. 

In this study, we propose \emph{historical credibility} that judges the credibility of reviews based on the historical ratings and textual reviews written by each reviewer. Fig.~1 shows actual examples of movie reviewers submitting indistinguishable reviews for multiple movies, which need to be considered untrustworthy. 
Fig.~1~(a) is an example where a movie reviewer has given the maximum point to all the movies they have reviewed. We indicate that indistinguishable overall positive reviews with similar expressions are written regardless of the movies.  
Fig.~1~(b) shows another example where a movie reviewer has given the minimum point and overall negative reviews to all the movies reviewed. Hence, we conclude, by evaluating their historical reviews, that we can identify movie reviewers that write indistinguishable reviews.

\begin{figure}[!h]
 \begin{subfigure}[t]{0.49\linewidth}
   \centering
   \includegraphics[width=.99\textwidth]{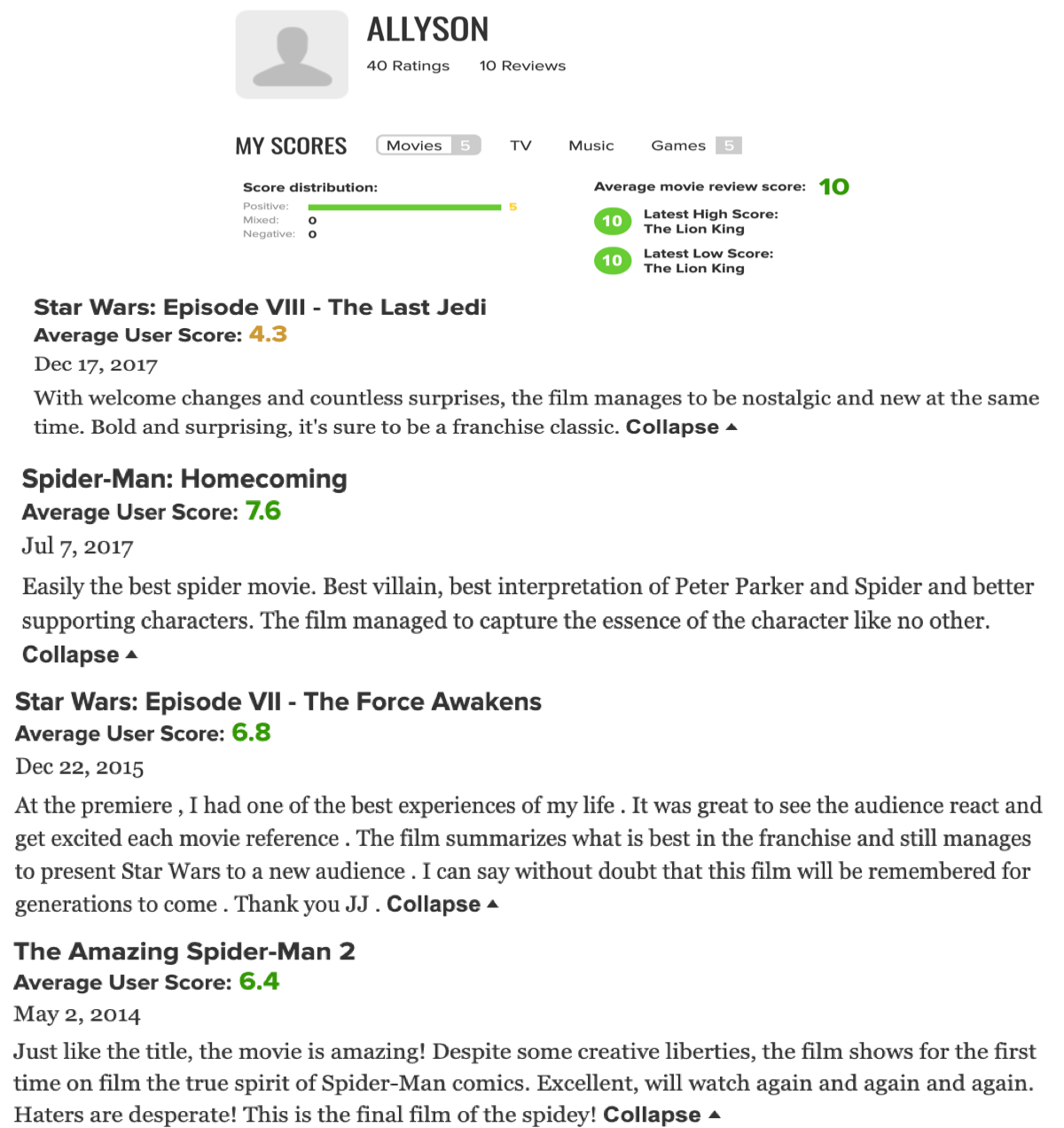}  
   \caption{A reviewer giving the maximum points.}
   \label{fig:fig2-a}
 \end{subfigure}
 \begin{subfigure}[t]{0.49\linewidth}
   \centering
   \includegraphics[width=.99\textwidth]{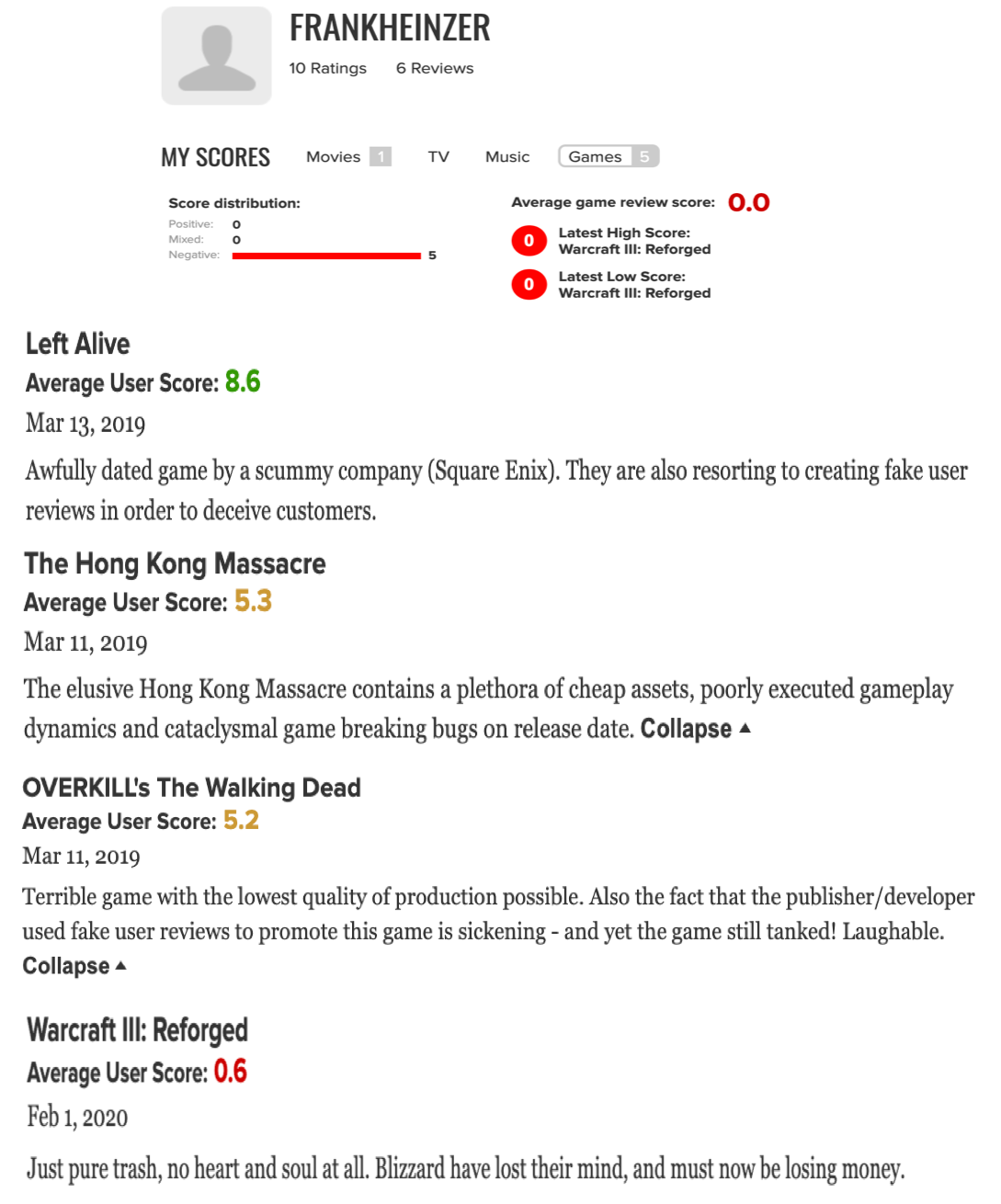}  
   \caption{A reviewer giving the minimum points.}
   \label{fig:fig2-b}
 \end{subfigure}%
 \caption{Examples of movie reviewers who write indistinguishable reviews.}
 \label{fig:fig2}
 \end{figure}

The contributions of the paper are summarized as follows: 

\begin{itemize}
    \item We propose historical credibility that judges the credibility of movie reviews based on the historical textual reviews and ratings written by each movie reviewer. For this, we present three kinds of criteria based on the ratings, sentiment scores of the textual reviews, and their correlation.
    \item We validate the effectiveness of the proposed historical credibility through extensive analysis. Specifically, we classify the entire reviews into the trusted and distrusted reviews by historical credibility and show that their characteristics are quite distinguishable in terms of three viewpoints: 1) distribution, 2) statistics, and 3) correlation. As a result, we obtain  evidence showing that historical credibility is effective. Specifically, the distribution and statistical analysis show that distrusted reviews tend to be more positive and indistinguishable than trusted reviews. Correlation analysis shows that the trusted reviews show more consistency in the ratings and reviews than the distrusted reviews. 
    \item We apply historical credibility to a weakly supervised model to classify a given review into trusted or distrusted one. First, it is significantly efficient because the entire data set is annotated according to the predefined criteria. Indeed, it can annotate 6,400 movie reviews only in 0.093 seconds, which occupies 0.55\% $\sim$ 1.88\% of the total learning time when we use LSTM and SVM as the learning model. This is a notable result because the usual human annotation takes much time and effort even for experts. Second, we show that the historical credibility-based classification model outperforms the textual review-based classification model. Specifically, the classification accuracy of the former outperforms that of the latter by up to 11.7\% $\sim$ 13.4\%. In addition, we clearly confirm that our classification model shows higher accuracy as the data size increases.
\end{itemize}

The paper is organized as follows. In Section~\ref{sec:sec2}, we describe related work. In Section~\ref{sec:sec4}, we propose historical credibility for judging the credibility of movie reviews. In Section~\ref{sec:validation}, we validate the effectiveness of the proposed historical credibility.
In Section~\ref{sec:sec5}, we apply historical credibility to a weakly supervised classification model and describe the experimental results to show its efficiency and accuracy. In Section~\ref{sec:sec6}, we further discuss the proposed historical credibility and its application to the classification model. In Section~\ref{sec:sec7}, we conclude the paper.
\section{Related Work}
\label{sec:sec2}

\subsection{Movie review analysis}

For movie reviews, most previous studies have focused on sentiment analysis of textual reviews, because the sentiment after watching the movie is an essential factor in analyzing the movie reviews~\cite{serrano2015sentiment}. Topal et al.~\cite{topal2016movie} constructed an emotion map according to emotions expressed after watching the movie and recommended movies to moviegoers according to this emotion map. Manek et al.~\cite{manek2017aspect} performed sentiment classification for movie reviews using a Gini index based on feature selection and a support vector machine (SVM) classifier. Chakraborty  et al.~\cite{chakraborty2018comparative} compared the performance of two clustering algorithms, namely $k$-means and $k$-means++, for emotion classification of movie reviews based on Word2Vec embedding of the textual review. Alsaqer et al.~\cite{alsaqer2017movie} improved the accuracy of movie review summarization based on sentiment analysis. He et al.~\cite{he2011self} proposed a self-training method using labeled features extracted from the existing sentiment lexicon on movie reviews. Elmurngi et al.~\cite{elmurngi2018fake} detected fake movie reviews based on sentiment analysis using supervised learning methods such as SVM and naïve Bayes. Sivakumar et al.~\cite{sivakumar2021analysis} presented a sentiment-based classification model on movie reviews by combining LSTM with word embedding to extract the semantics between the neighboring words. Dashtipour et al.~\cite{dashtipour2021sentiment} applied CNN and LSTM to analyze the sentiment of the movie reviews.

\subsection{Credibility on online reviews} 

To the best of our knowledge, there have been no public methods that consider the credibility of movie reviews. Many well-known existing movie review management systems, including Metacritic and Rotten Tomatoes, do not consider movie review credibility in calculating the movie's overall rating. IMDb is known to exclude distrusted reviews in calculating the movie's overall rating; however, since the specific methods are not disclosed, we cannot evaluate their effectiveness. 

On the other hand, many research efforts have been invested in evaluating the credibility of online reviews. It is well known that judging the credibility of online reviews is important because they affect consumers' decision-making~\cite{lee2011helpful, mudambi2010makes, ham2019subjective}. Liu et al.~\cite{liu2015makes} evaluated the effectiveness of reviews of travel products based on two aspects: 1) reviewer-related information (e.g., user profiles) and 2) review-related information (i.e., ratings and length of reviews). Ghose et al.~\cite{ghose2010estimating} explored multiple aspects of textual reviews, namely, subjectivity levels, readability, and spelling errors, to understand the helpfulness of reviews in terms of economic and social outcomes. Hochmeister et al.~\cite{hochmeister2013destination} compared destination experts (i.e., the most active members) in TripAdvisor with normal reviewers and showed that destination experts have more influence in the TripAdvisor community. Barbado et al.~\cite{barbado2019framework} proposed a method for detecting fake reviews in the domain of consumer electronics businesses (i.e., Yelp businesses), labeling reviews as trustworthy or fake. Reyes et al.~\cite{reyes2019impact} studied five chosen factors on reviews that can affect the decision-making of customers by exploring TripAdvisor reviews. Luo et al.~\cite{luo2021makes} defined factors influencing review's helpfulness from the perspectives of both reviews and reviewers. Aghakhani et al.~\cite{aghakhani2021online} defined variables in terms of consistency and inconsistency of reviews and ratings to evaluate the review's helpfulness. 

Many previous studies have shown that the helpfulness vote by other reviewers is a critical measure for evaluating review credibility~\cite{fang2016analysis, shan2016credible, yang2017exploring, ren2019examining, ahmad2015expressed, hong2017understanding}. Fang et al.~\cite{fang2016analysis} used the helpfulness vote as the criterion for determining review credibility to explore factors affecting the credibility of reviews. Shan et al.~\cite{shan2016credible} showed that the reviewer reputation generated by peer ratings has a great influence on credibility evaluation. Yang et al.~\cite{yang2017exploring} examined six factors (specifically, reviewer locations, reviewer levels, helpfulness votes, review ratings, review lengths, and review photos) and showed that review ratings and helpfulness votes were the most influential factors. Gang et al.~\cite{ren2019examining} examined the relationship between emotions embedded in online reviews and review helpfulness and analyzed different effects of each emotion on perceived review helpfulness. Jabr et al.~\cite{jabr2021review} measured the credibility of product reviews based on helpfulness votes.

\subsection{Weakly supervised learning} 

Weakly supervised learning allows us to annotate large-scale data at a very fast rate, at the cost of potentially incurring some inaccurate annotation~\cite{zhou2018brief}. 
Typical annotation methods for weakly supervised learning can be classified into three categories~\cite{dong2018data}: 1) crowdsourcing, 2) distant supervision, and 3) labeling functions. Crowdsourcing is a form of supervised learning, but annotation of data is performed by several annotators who are not experts~\cite{raykar2010learning}. Distant supervision extracts the structure from a data set using large-scale existing databases; the extracted structure is used for annotation~\cite{mintz2009distant}. Labeling functions are constructed from defined rules for annotating data~\cite{ratner2017snorkel, ratner2016data}. Recently, weakly supervised learning has been adopted in many studies to resolve the difficulty of data annotation. For instance, Lee et al.~\cite{lee2018sentiment} applied weakly supervised learning to identify keywords for classifying positive and negative sentences. Lin et al.~\cite{lin2018multi} used weakly supervised learning for understanding the users' sentiments from social media contents containing different types of data such as texts and images. Taher et al.~\cite{taher2021adversarial} labeled a data set with weakly supervised sentiment tags using a sentiment vocabulary network. In this study, we also apply the predefined criteria defined based on historical credibility for weakly supervised learning, showing the fast annotation and effectiveness.

\section{Historical Credibility}
\label{sec:sec4}

In this section, we propose new criteria for judging the credibility of the movie reviewers based on their historical reviews. In Section~\ref{sec:sec4-11}, we explain the motivation for devising historical credibility. In Section~\ref{sec:sec4-22}, we define historical credibility based on three kinds of features: 1) ratings, 2) sentiment scores, and 3) their correlation.

\subsection{Motivation}
\label{sec:sec4-11}

In this section, we discuss the necessity of devising new criteria to determine the credibility of the movie review. To evaluate the credibility of reviews, Weng et al.~\cite{weng2019cats} identified the differences between the reviews on fraud items and those on normal items on the e-commerce by comparing the distribution according to sentiment analysis. Fig.~\ref{fig:fig2}~(a) shows that the distribution of sentiment scores for fraud items and that of normal items on e-commerce are quite different. That is, the sentiment distribution of the reviews on fraud items is long-tailed, with very few negative sentiment reviews and an extremely positive tendency while normal items have a relatively uniform distribution. 

\begin{figure}[!h]
 \begin{subfigure}[t]{0.49\linewidth}
   \centering
   \includegraphics[width=0.99\textwidth,height=0.17\textheight]{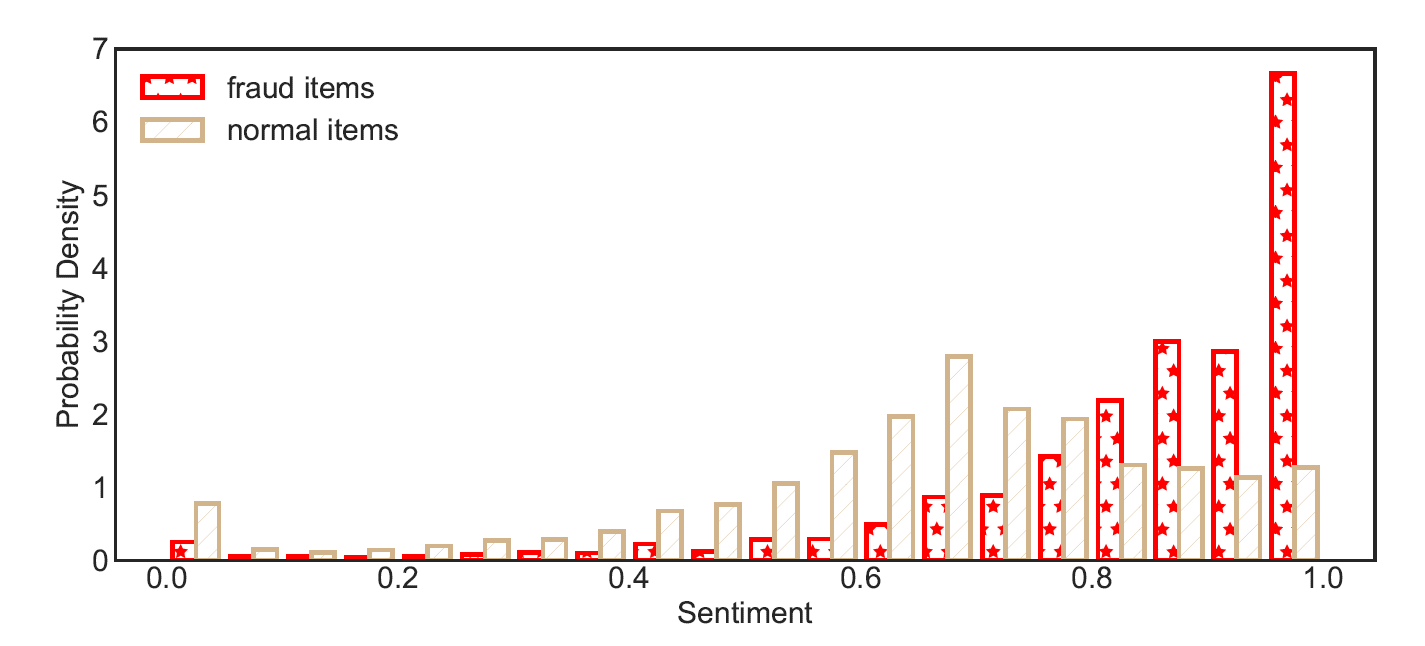}
   \caption{Fraud and normal items' comments on e-commerce ~\cite{weng2019cats}.}
   \label{fig:fig2-a}
 \end{subfigure}
 \begin{subfigure}[t]{0.49\linewidth}
   \centering
   \includegraphics[width=0.99\textwidth,height=0.16\textheight]{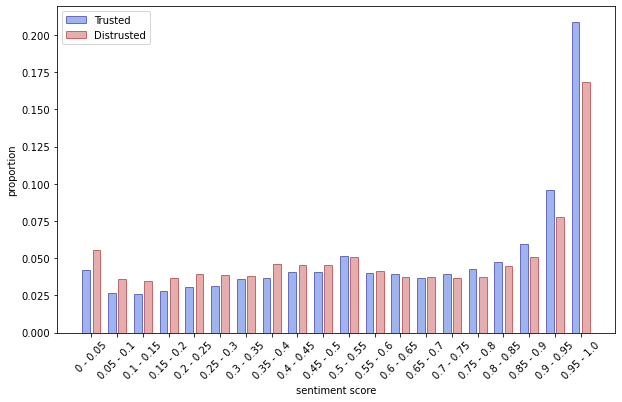}
   \caption{Trusted and distrusted movie reviews classified by helpfulness votes.}
   \label{fig:fig2-b}
 \end{subfigure}%
 \caption{Comparison of sentiment distribution between comments on e-commerce and movie reviews.}
 \label{fig:fig2}
 \end{figure}

The helpfulness vote has been used as a meaningful measure for evaluating the reviews. 
Shan et al.~\cite{shan2016credible} and Yang et al.~\cite{yang2017exploring} showed that the helpfulness vote has a great influence on credibility evaluation. Fang et al.~\cite{fang2016analysis} also used the helpfulness vote as a criterion to evaluate the influence of various factors on review credibility. We also consider the helpfulness vote as an effective criterion to judge the credibility of movie reviews. We define each movie review as trusted when the number of helpfulness votes for the review exceeds the number of unhelpfulness votes. Fig.~\ref{fig:fig2}~(b) shows the sentiment distribution of trusted and distrusted reviews classified by the helpfulness vote with the same approach in Fig.~\ref{fig:fig2}~(a). However, we indicate that the overall distribution of Fig.~\ref{fig:fig2}~(b) is quite different from Fig.~\ref{fig:fig2}~(a). Hence, the positive tendency of the distrusted reviews in Fig.~\ref{fig:fig2}~(b) is not high as in Fig.~\ref{fig:fig2}~(a). In addition, the proportion of the distrusted reviews with low sentiment scores is much higher than that in Fig.~\ref{fig:fig2}~(a). Therefore, we conclude that the helpfulness vote is not that effective to evaluate the credibility of the movie reviews.

\subsection{Three Criteria for Historical Credibility}
\label{sec:sec4-22}
 
In this study, we define criteria that can classify movie reviews according to their credibility by utilizing the historical reviews of movie reviewers. Our basic idea is to identify trusted or distrusted movie reviewers according to the discrimination of historical reviews written by them. Chen et al.~\cite{chen2001computing} showed that online reviewers with high reputations discriminate against the objects, indicating that the standard deviation of the overall rating increases. 
In this respect, we focus on the fact that credible reviewers write distinguishable reviews and ratings depending on the movie, i.e., their distribution becomes variable. For this, we define two different kinds of criteria to measure the reviewer’s credibility according to the degree of divergence of the historical reviews: 1) historical rating and 2) historical sentiment. Last, we define a criterion according to the coincidence between them: historical correlation.

\begin{definition}[Historical Rating-based Credibility]
We use the standard deviation of the historical ratings of each reviewer to measure the credibility. We define the upper ranked reviewers according to the  standard deviation of written ratings as \emph{trusted reviewers} and the reviews written by them as \emph{trusted reviews}; We define the remaining reviewers as \emph{distrusted reviewers} and reviews written by them as \emph{distrusted reviews}. \hfill
\end{definition}

\begin{definition}[Historical Sentiment-based Credibility]
We use the standard deviation of the historical sentiment scores of the textual reviews of each reviewer to measure the credibility. 
We define the upper ranked reviewers according to the standard deviation of sentiment scores of written reviews as \emph{trusted reviewers} and the reviews written by them as \emph{trusted reviews}; We define the remaining reviewers as \emph{distrusted reviewers} and reviews written by them as \emph{distrusted reviews}. \hfill
\end{definition}

\begin{definition}[Historical Correlation-based Credibility]
We use Spearman’s Correlation Coefficient to measure the relevance of a set of historical sentiment scores of the textual reviews and a set of historical ratings of each reviewer. We define the upper ranked reviewers according to the correlation coefficient as \emph{trusted reviewers} and the reviews written by them as \emph{trusted reviews}; We define the remaining reviewers as \emph{distrusted reviewers} and reviews written by them as \emph{distrusted reviews}. \hfill
\end{definition}


We elaborate on the value of the proposed historical credibility in terms of two viewpoints. First, it is inherently difficult to separate trusted and distrusted reviews only from the reviews consisting of short texts. Historical credibility provides clear criteria to classify them. As a result, we can apply these clear criteria to a classification model based on weakly supervised learning in Section 5. Second, historical credibility provides an efficient and automatic evaluation of helpfulness to the reviews compared to the helpfulness vote. Although the helpfulness vote is considered an effective way to judge the credibility of the reviews, it has the following limitations: 1) it needs to be performed manually and 2) it requires enough time to label them, and hence, recent reviews cannot be labeled. 


\section{Validation}
\label{sec:validation}


In this section, we validate the proposed historical credibility compared to the helpfulness vote. In Section 4.1, we present the overall framework for the validation. For trusted or distributed reviews classified by historical credibility, we show distribution analysis, statistical analysis, and correlation analysis in Sections 4.2, 4.3, and 4.4, respectively.

\subsection{Overall Framework}

\begin{figure}[!ht]
  \centering
  \includegraphics[width=0.99\linewidth]{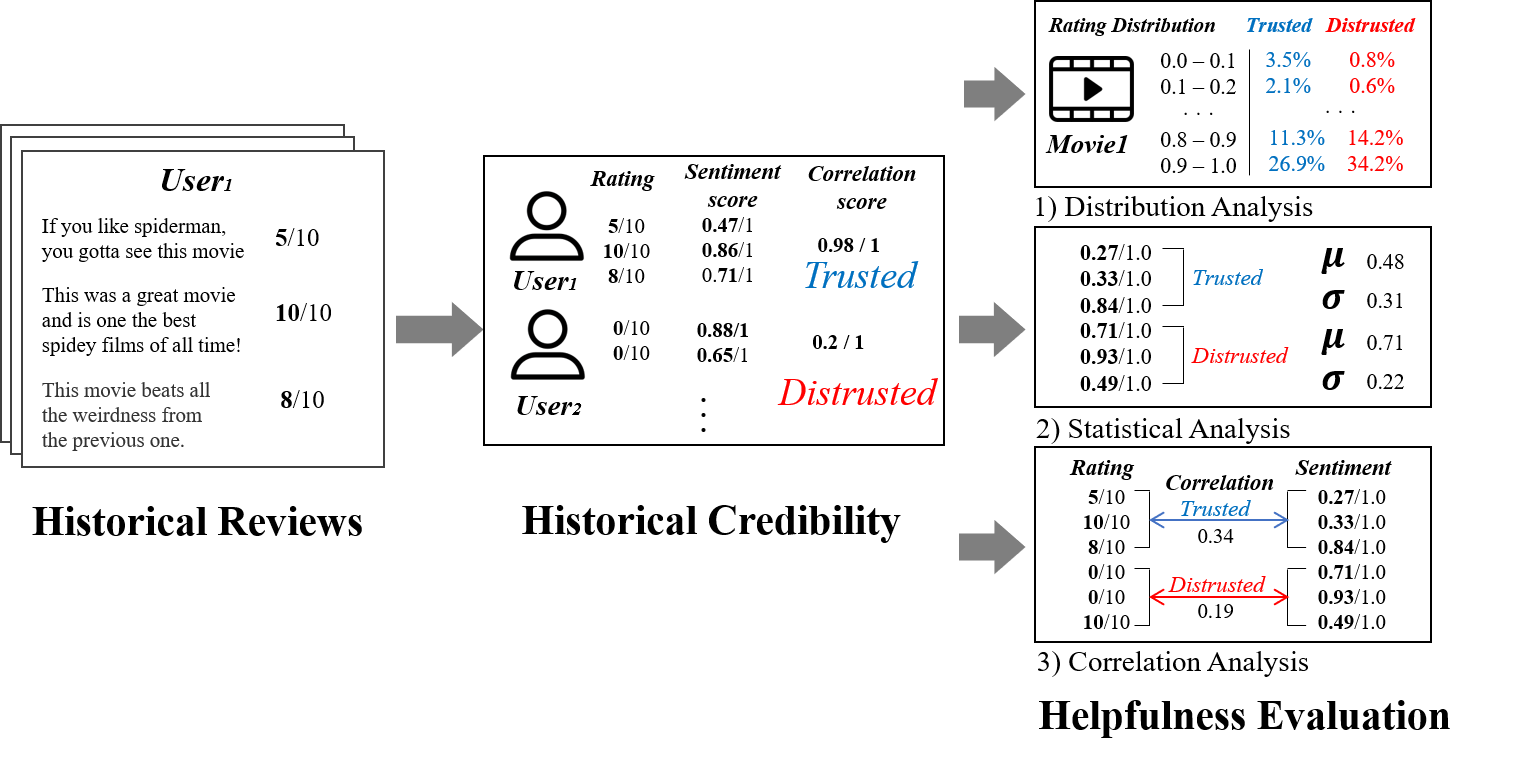}  
  \caption{A framework for validating the effectiveness of historical credibility.}
  \label{fig:fig55-a}
\end{figure}

In this study, we validate the effectiveness of the proposed historical credibility. Fig.~\ref{fig:fig55-a} shows a framework for validating the effectiveness of historical credibility. First, from the historical reviews of the reviewers, we classify the entire reviews according to three criteria for historical credibility to trusted or distrusted reviews. Then, we perform distribution, statistical, and correlation analyses for trusted and distrusted reviews. For this validation, we used movie review data sets collected from Naver Movie, the largest Korean movie review website, and released them at Mendeley Data\footnote{http://dx.doi.org/10.17632/jb5knzh8yv.6}. Table~\ref{tab:tab19} shows their statistics by the movie. It shows the number of total reviews, that of trusted and distrusted reviews classified by helpfulness votes. For the analysis of historical reviews for each reviewer, we only consider reviewers with three or more past reviews. In addition, for the fair comparison with the helpfulness vote, we define the same number of trusted and distrusted reviews for each proposed criterion with that of the helpfulness vote by the movie.

\begin{table}
\fontsize{9}{10}\selectfont
\center{
\caption{Movie review data sets for the validation.} 
\begin{tabular}{|c|c|c|c|}

\hline
\textbf{Movies} & \textbf{Genre} & \textbf{Total reviews}  & \textbf{\begin{tabular}[c]{@{}c@{}}Trusted / distrusted reviews \end{tabular}} \\
\hline
\hline Assassination & \makecell{Action,\\ Drama} & 11,298 & 6,821 / 4,477 \\
\hline \makecell{Confidential\\ Assignment} & \makecell{Action} & 11,526 & 6,504 / 5,022 \\
\hline 1987 & \makecell{Drama} & 15,416 & 11,926 / 3,490\\
\hline A taxi driver & \makecell{Drama, \\Family} & 19,905 & 14,976 / 4,929 \\
\hline \makecell{Intimate\\ strangers} & \makecell{Drama, \\Comedy} & 9,456 & 4,571 / 4,885 \\
\hline \makecell{The Spy \\Gone North} & \makecell{Drama} & 12,464 & 6,703 / 5,761 \\
\hline The outlaws & \makecell{Action} & 16,327 & 14,582 / 1,745 \\
\hline
\hline \textbf{Total} & \makecell{} & 96,392 & 66,083 / 30,309 \\
\hline
\end{tabular}
\label{tab:tab19}
}
\end{table}

\subsection{Distribution Analysis}

We adopt the approach in Weng et al.~\cite{weng2019cats} that analyzes the distribution based on sentiment analysis to the three definitions proposed in this paper: 1) ratings, 2) sentiment score for the textual reviews, and 3) correlation coefficient between them. For each definition, we classify the entire review into two groups, i.e., trusted and distrusted reviews according to historical credibility, and conduct distribution analysis for each group. To measure the sentiment score of the movie review, we extract the top 200 tokens with the high frequency using KoNLPy\footnote{https://konlpy.org/en/latest/} and NLTK\footnote{https://www.nltk.org/} libraries and create a pre-trained emotional model through the count vectorization. For the training data set, we used Naver movie review data sets\footnote{https://github.com/e9t/nsmc/} and, for the testing data set, we used seven movie reviews in Table 2, to completely separate both of them. We define reviews less than 4 points as the negative reviews and reviews higher than 7 points as the positive reviews according to the existing data annotation method~\cite{maas2011learning}.

\begin{figure}[!h]
\begin{subfigure}[a]{0.5\linewidth}
  \centering
  \includegraphics[width=0.97\linewidth]{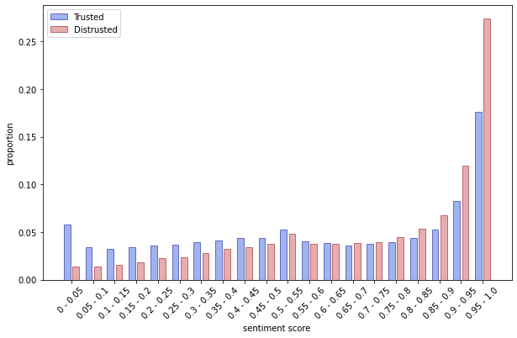}  
  \caption{Definition~1.}
  \label{fig:fig3a}
\end{subfigure}
\begin{subfigure}[c]{0.5\linewidth}
  \centering
  \includegraphics[width=0.97\linewidth]{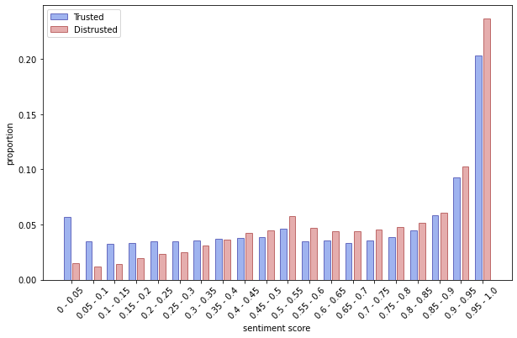}  
  \caption{Definition~2.}
  \label{fig:fig3b}
\end{subfigure}
\begin{subfigure}[c]{0.5\linewidth}
  \centering
  \includegraphics[width=0.97\linewidth]{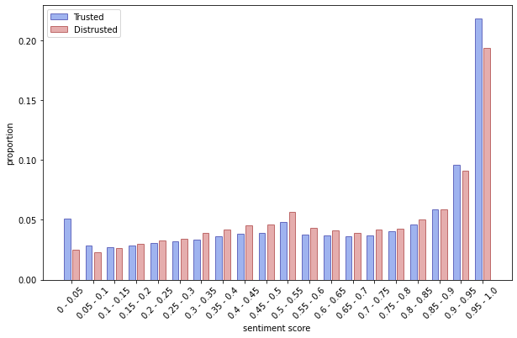}  
  \caption{Definition~3.}
  \label{fig:fig3c}
\end{subfigure}
\begin{subfigure}[d]{0.5\linewidth}
  \centering
  \includegraphics[width=0.97\linewidth]{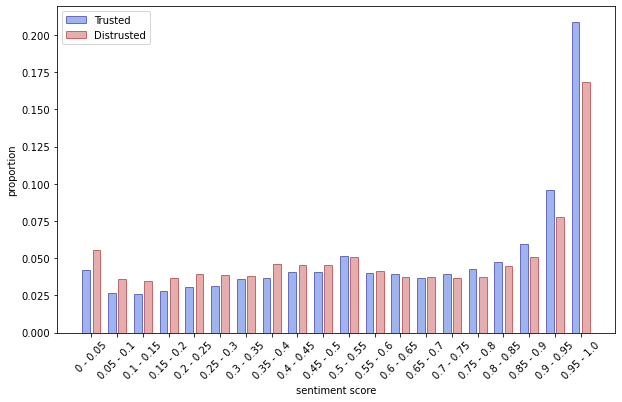}  
  \caption{Helpfulness votes.}
  \label{fig:fig3d}
\end{subfigure}
\caption{Distribution of textual reviews’ sentiment scores defined by historical credibility and helpfulness vote.}
\label{fig:fig12}
\end{figure}

Fig.~\ref{fig:fig12} compares the distribution of reviews’ sentiment scores for the trusted and distrusted reviews defined by historical credibility and helpfulness vote. We normalize the sentiment score as the value from 0 to 1 where negative reviews are close to 0 and positive reviews are to 1. As shown in Fig.~\ref{fig:fig12}, our historical credibility shows similar trends to the results presented in Weng et al.~\cite{weng2019cats}. That is, the extremely positive tendency increases and the extremely negative tendency decreases in the case of distrusted reviews while relatively even distribution is observed in the case of trusted reviews. Specifically, Definition~1 shows that the portion between 0.95 and 1.0 in the sentiment score for distrusted reviews is greater than that for trusted reviews by 55.99\%; the portion between 0.0 and 0.05 in the sentiment score for distrusted reviews is lower than that for trusted reviews by 77.11\%. Definition~2 shows the overall similar trends to Definition~1. 
In the case of Definition~3, the extremely negative tendency does also decrease in the case of distrusted reviews. However, we indicate that the portion of the extremely positive sentiments in the distrusted reviews is smaller than that of trusted reviews unlike in Definitions 1 and 2. Here, we analyze that, unlike Definitions 1 and 2 considering the ratings and sentiment scores separately, when we consider both of them together, they are highly consistent in the trusted reviews in both extremely positive and negative cases, which has a greater influence on the results.

On the other hand, the helpfulness vote in Fig.~\ref{fig:fig12}~(d) shows a completely different trend to historical credibility. That is, distrusted reviews show a larger portion for the negative reviews and a smaller portion for the positive reviews compared to the trusted reviews. 

\begin{figure}[!h]
\begin{subfigure}[a]{0.5\linewidth}
  \centering
  \includegraphics[width=0.97\linewidth]{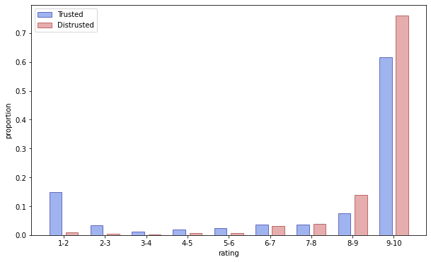}  
  \caption{Definition~1}
  \label{fig:fig32a}
\end{subfigure}
\begin{subfigure}[c]{0.5\linewidth}
  \centering
  \includegraphics[width=0.97\linewidth]{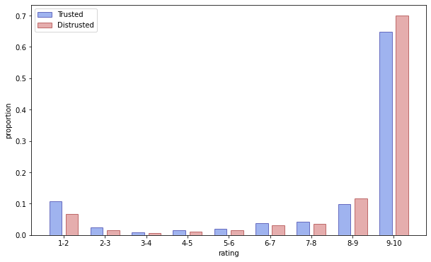}  
  \caption{Definition~2}
  \label{fig:fig32b}
\end{subfigure}
\begin{subfigure}[c]{0.5\linewidth}
  \centering
  \includegraphics[width=0.97\linewidth]{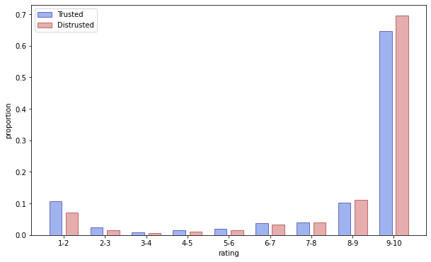}  
  \caption{Definition~3}
  \label{fig:fig32c}
\end{subfigure}
\begin{subfigure}[d]{0.5\linewidth}
  \centering
  \includegraphics[width=0.97\linewidth]{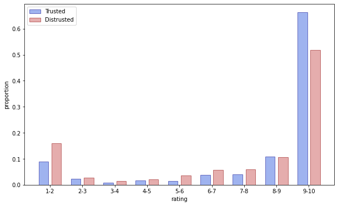}  
  \caption{Helpfulness votes}
  \label{fig:fig32d}
\end{subfigure}
\caption{Distribution of reviews’ ratings for the reviews defined by historical credibility and helpfulness vote.}
\label{fig:fig13}
\end{figure}

Fig.~\ref{fig:fig13} compares the distribution of reviews’ ratings for the trusted and distrusted reviews defined by the historical credibility and helpfulness vote. The overall trends are similar to the results in ~\ref{fig:fig12} and all the definitions for the historical credibility show the consistent results. Specifically, Definition~1, shows that the portion between 9 and 10 in the rating for distrusted reviews is greater than that for trusted reviews by 23.29\%; the portion between 0 and 1 in the rating for distrusted reviews is lower than that for trusted reviews by 93.93\%. Both Definitions~2 and 3 also show similar trends to Definition~1. 
On the other hand, the helpfulness vote in Fig.~\ref{fig:fig13}~(d) shows the opposite result in both extremely positive and negative cases, similar to Fig.~\ref{fig:fig12}~(d). 

\subsection{Statistical Analysis}

\begin{figure}[!h]
\begin{subfigure}[a]{0.5\linewidth}
  \centering
  \includegraphics[width=0.97\linewidth]{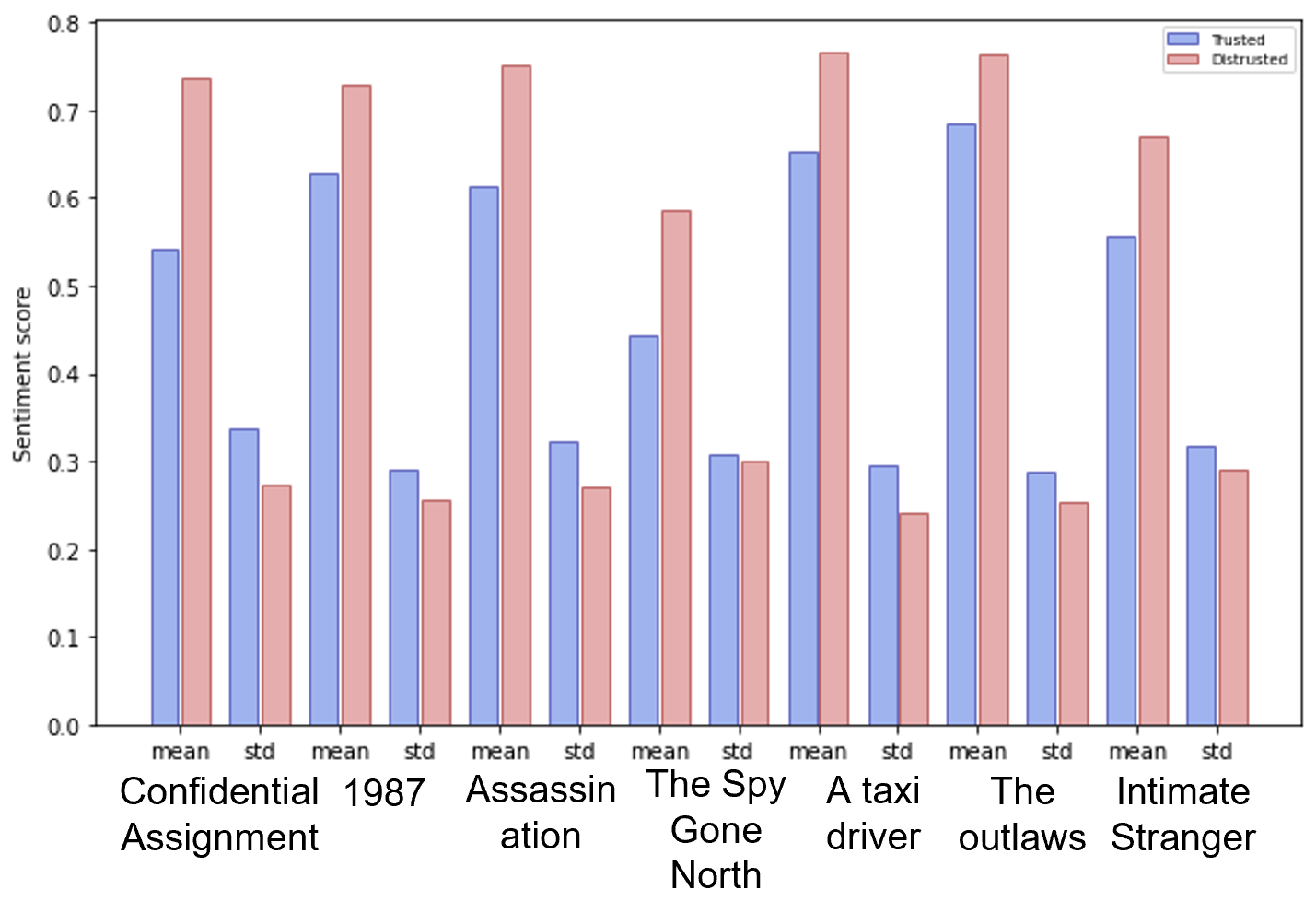}  
  \caption{Definition~1.}
  \label{fig:fig41a}
\end{subfigure}
\begin{subfigure}[c]{0.5\linewidth}
  \centering
  \includegraphics[width=0.97\linewidth]{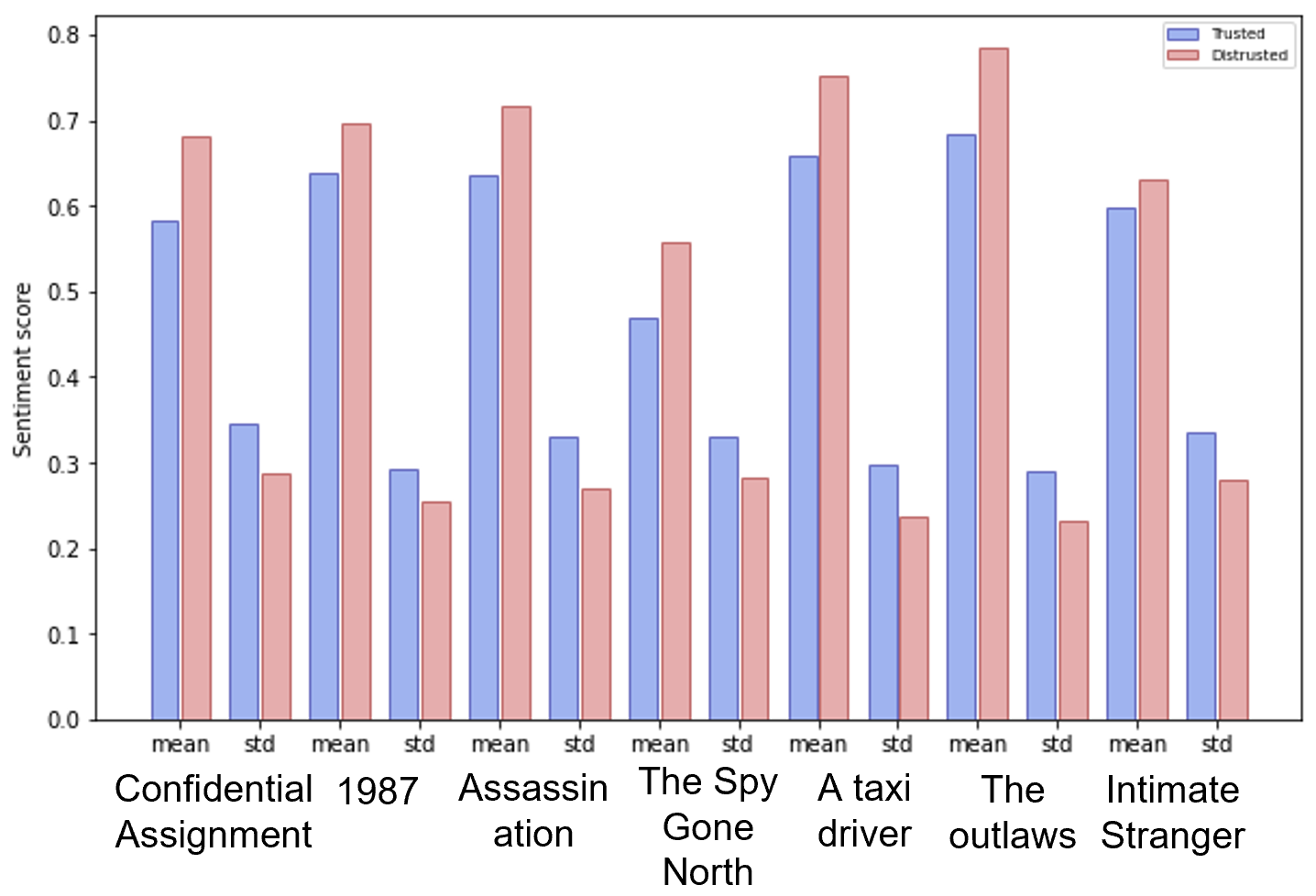}  
  \caption{Definition~2.}
  \label{fig:fig41b}
\end{subfigure}
\begin{subfigure}[c]{0.5\linewidth}
  \centering
  \includegraphics[width=0.97\linewidth]{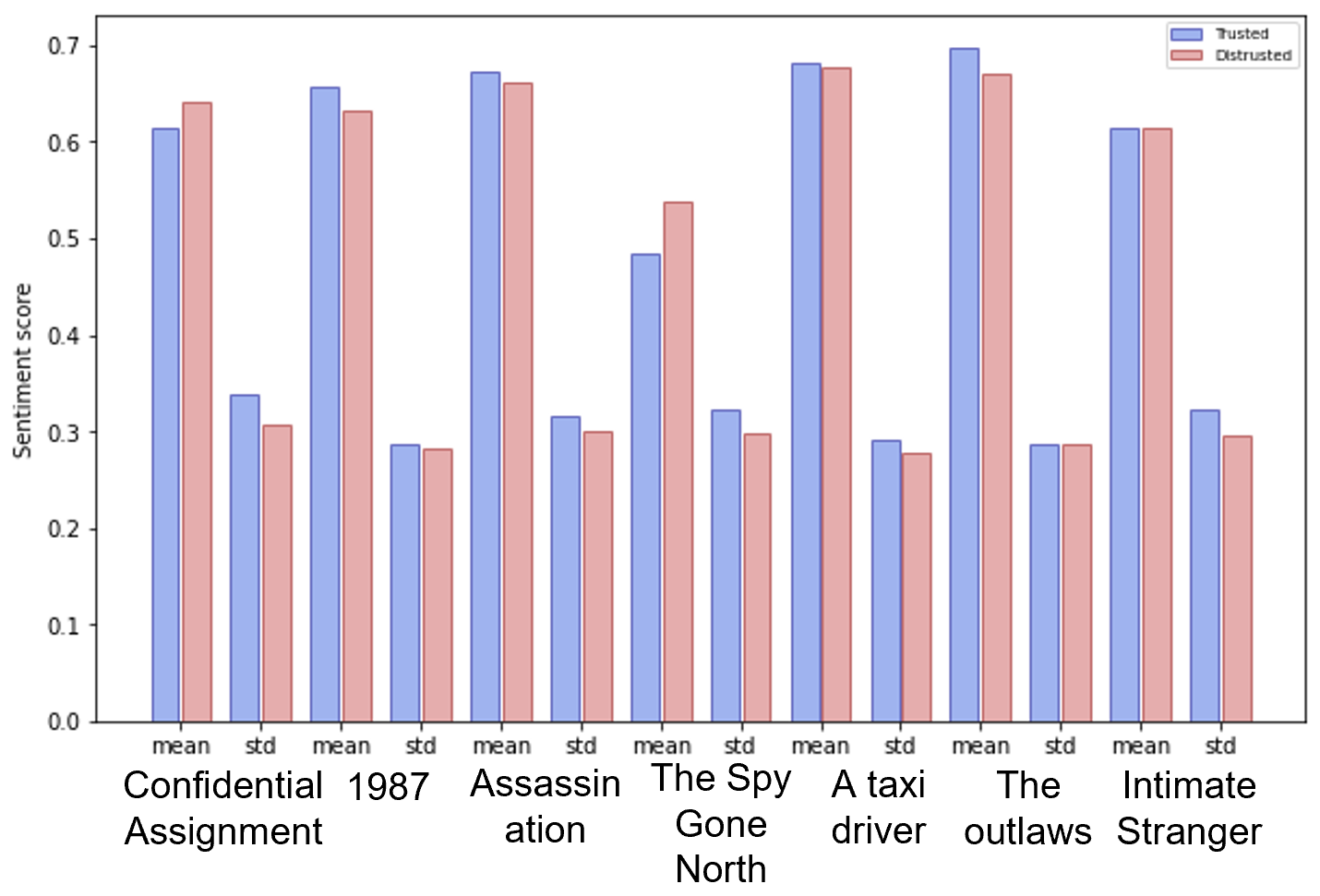}  
  \caption{Definition~3.}
  \label{fig:fig41c}
\end{subfigure}
\begin{subfigure}[d]{0.5\linewidth}
  \centering
  \includegraphics[width=0.97\linewidth]{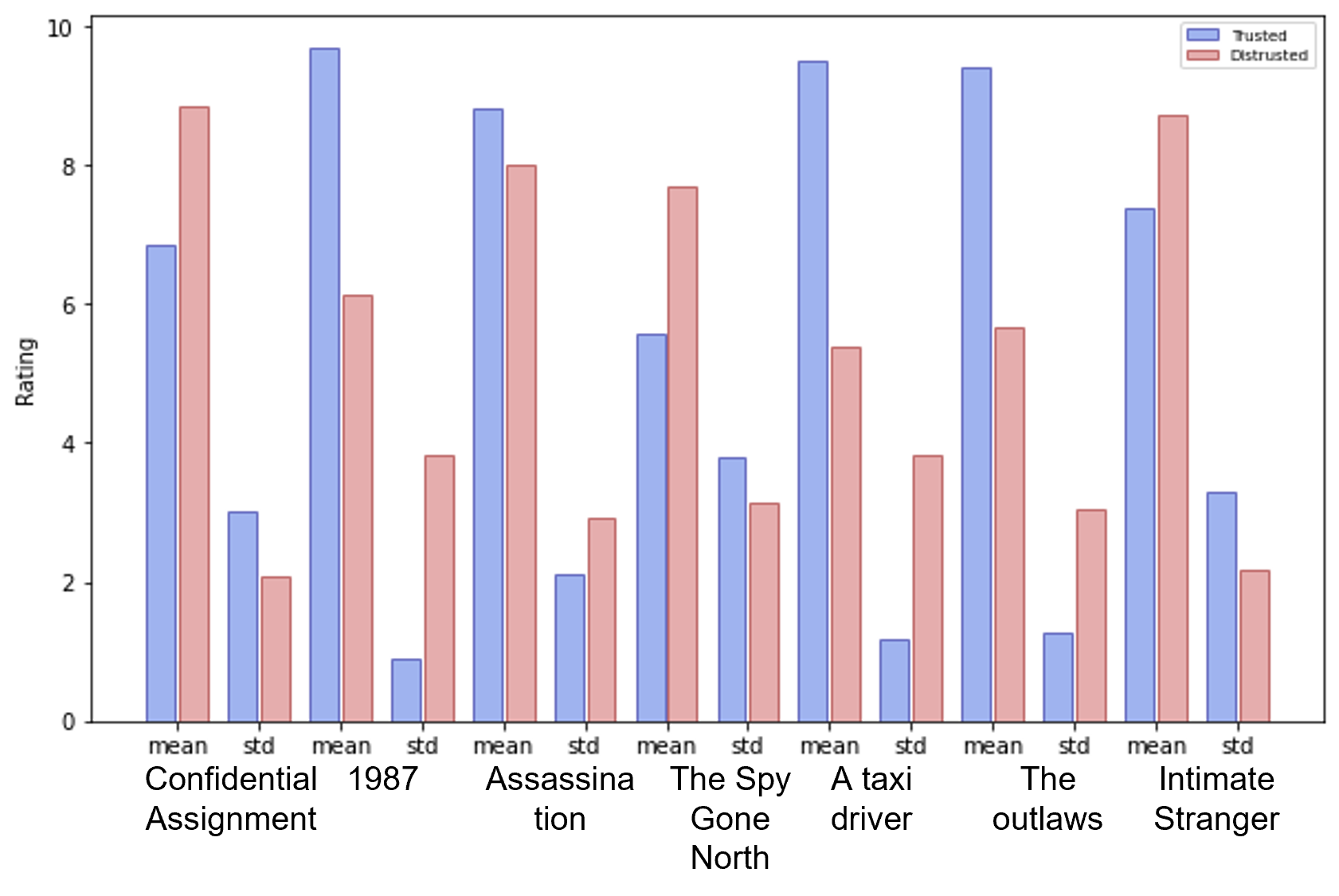}  
  \caption{Helpfulness votes.}
  \label{fig:fig41d}
\end{subfigure}
\caption{Statistical analysis of reviews' sentiment scores for trusted and distrusted reviews defined by historical credibility and helpfulness vote.}
\label{fig:fig14}
\end{figure}

Fig.~\ref{fig:fig14} compares the mean and standard deviation of the sentiment scores for trusted and distrusted reviews defined by historical credibility and the helpfulness vote. The result indicates that the trusted reviews show a lower mean and higher standard deviation compared to the distrusted reviews, which shows that trusted reviews write more distinguishable reviews than distrusted reviewers. This shows a consistent result with the claim by Chen et al.~\cite{chen2001computing}. Specifically, the mean of the trusted reviews defined by Definition~1 is 0.6126 and that of distrusted reviews is 0.7048; the mean of the trusted reviews by Definition~2 is 0.6267 and that of distrusted reviews is 0.6742. In contrast, in the case of Definition~3, the mean of the trusted reviews is similar to that of distrusted reviewers. Here again, extremely consistent tendency of the trusted reviews for positive reviews by Definition~3 in Fig.~\ref{fig:fig12}~(c) also affects the results. In addition, the standard deviation of trusted reviews defined by Definition~1 is 0.3117 and that of distrusted reviews is 0.2804; the standard deviation of trusted reviews by Definition~2 is 0.3160 and that of distrusted reviews is 0.2774; the standard deviation of trusted reviews by Definition~3 is 0.3098 and that of distrusted reviews is 0.2948. In contrast, Fig.~\ref{fig:fig14}~(d) shows that the reviews classified by the helpfulness vote do not show clear trends on overall mean and standard deviation of the trusted and distrusted reviews.

\begin{figure}[!h]
\begin{subfigure}[a]{0.5\linewidth}
  \centering
  \includegraphics[width=0.97\linewidth]{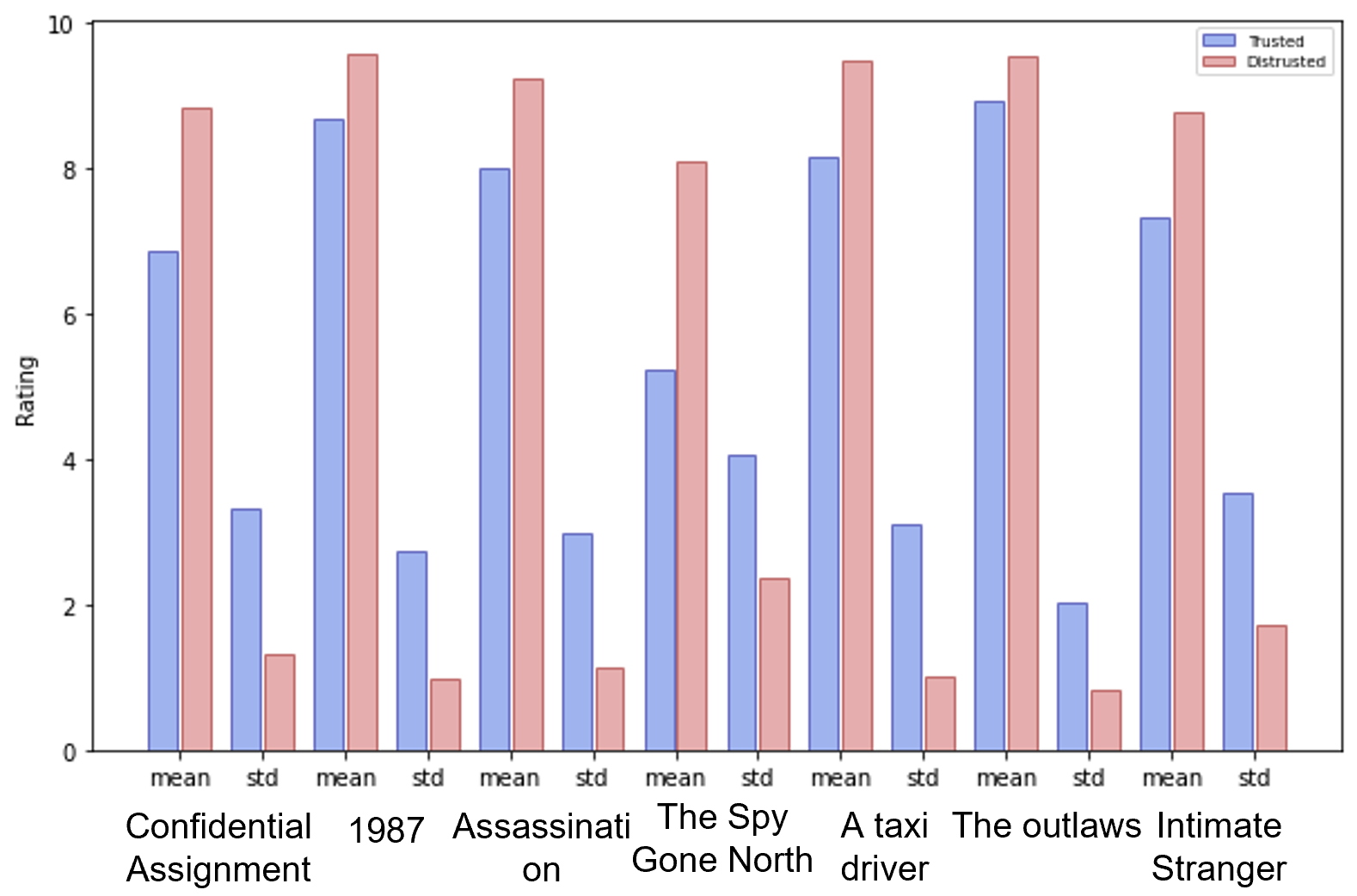}  
  \caption{Definition~1.}
  \label{fig:fig42a}
\end{subfigure}
\begin{subfigure}[c]{0.5\linewidth}
  \centering
  \includegraphics[width=0.97\linewidth]{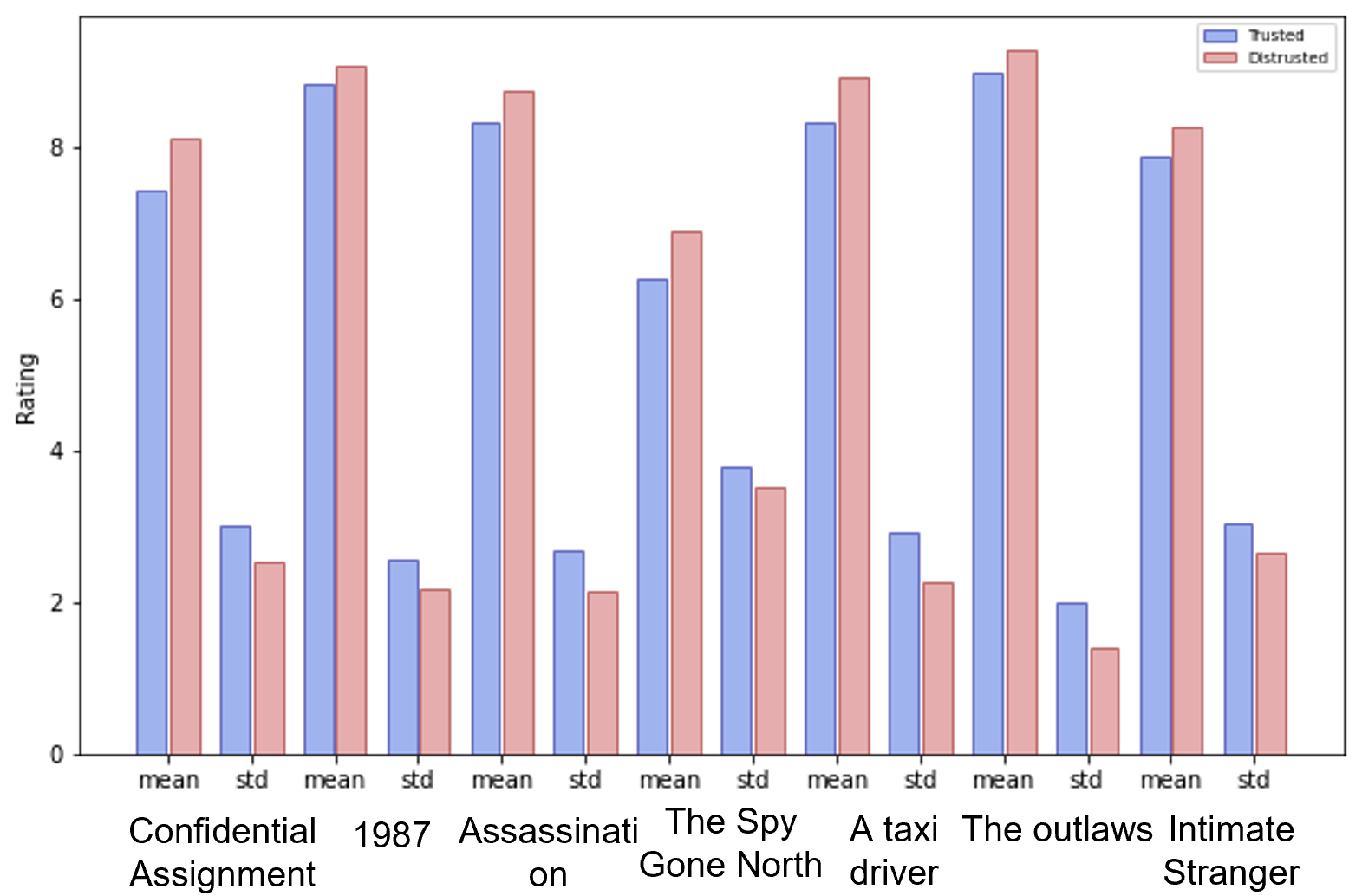}  
  \caption{Definition~2.}
  \label{fig:fig42b}
\end{subfigure}
\begin{subfigure}[c]{0.5\linewidth}
  \centering
  \includegraphics[width=0.97\linewidth]{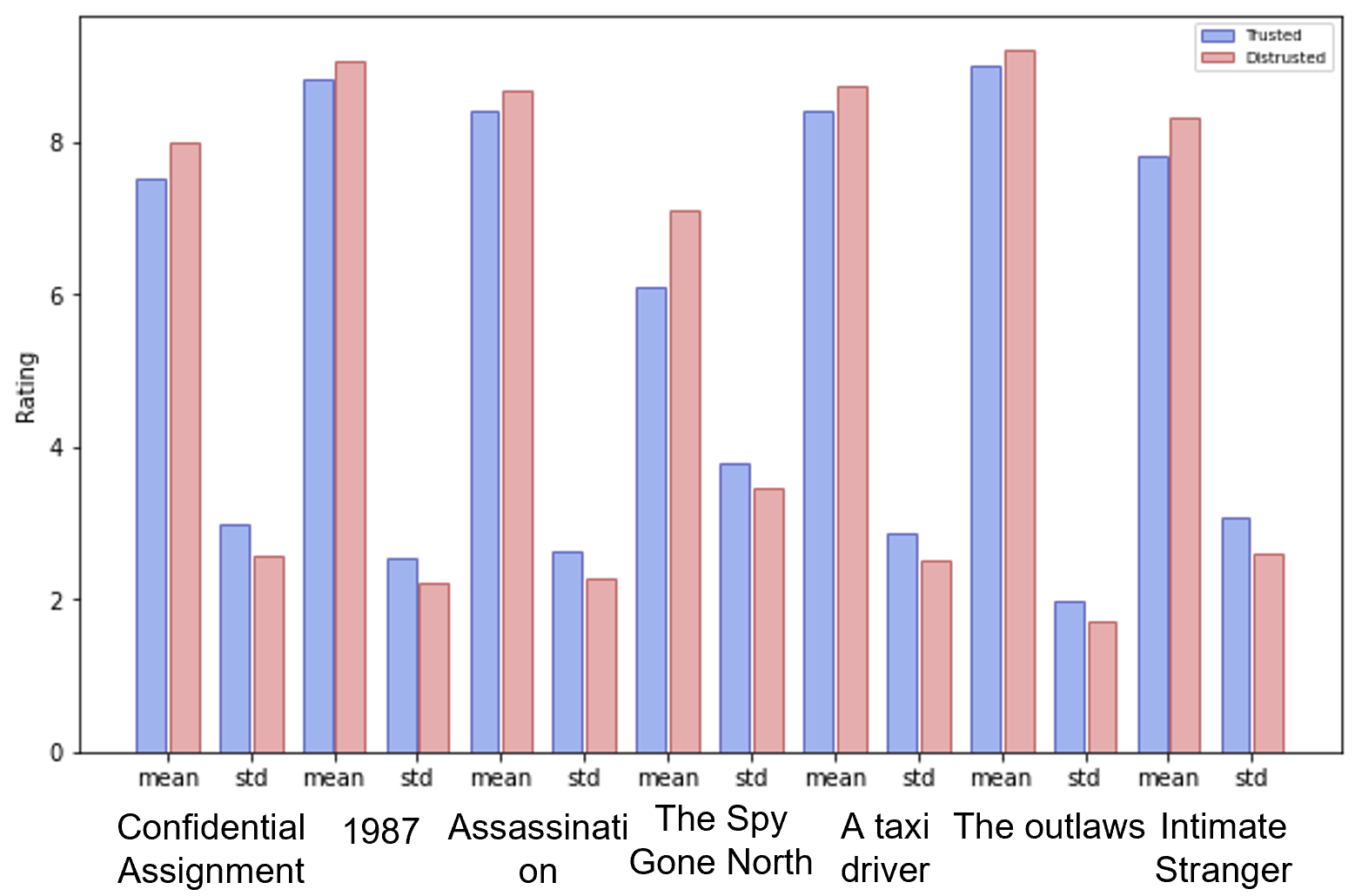}  
  \caption{Definition~3.}
  \label{fig:fig42c}
\end{subfigure}
\begin{subfigure}[d]{0.5\linewidth}
  \centering
  \includegraphics[width=0.97\linewidth]{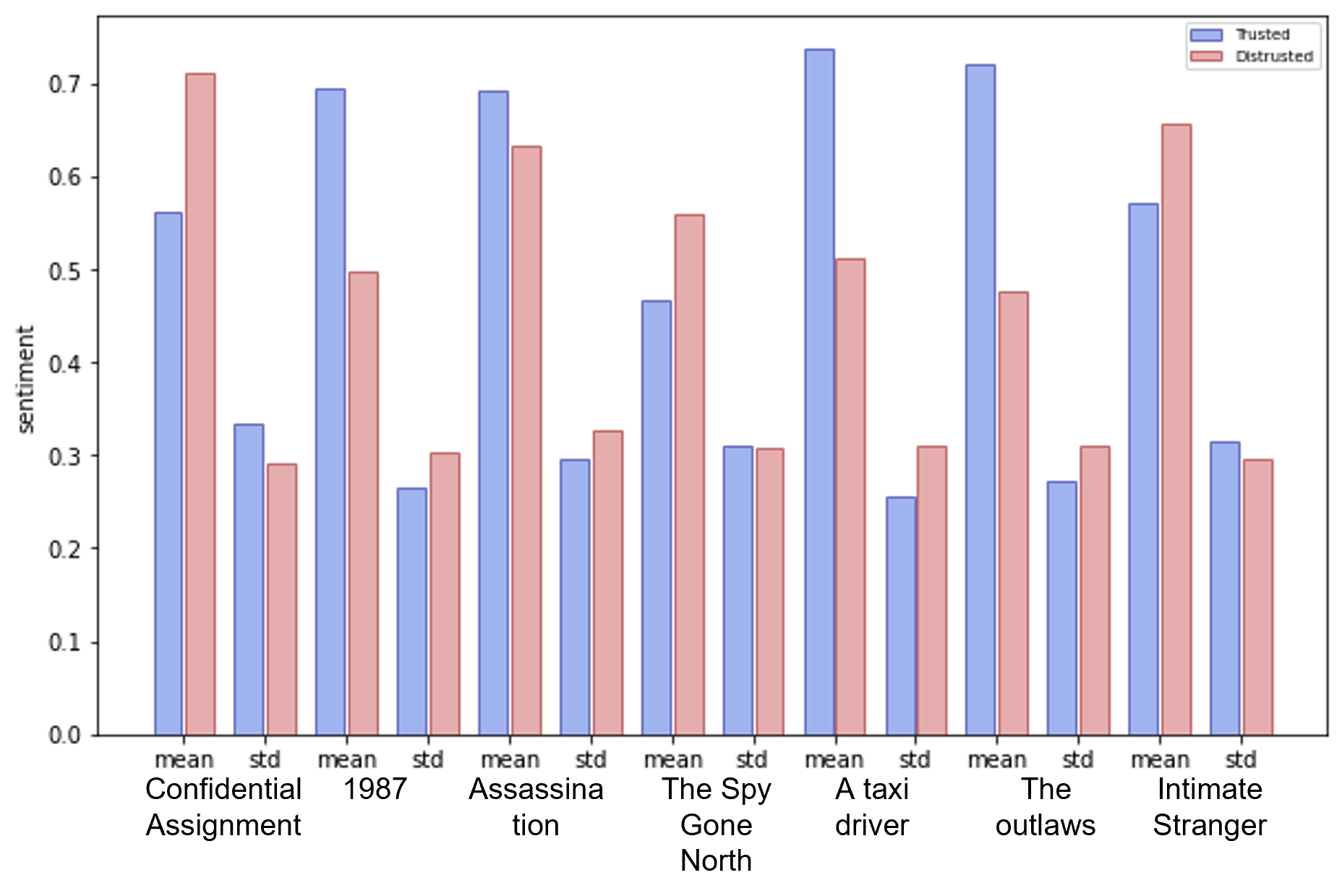}  
  \caption{Helpfulness votes.}
  \label{fig:fig42d}
\end{subfigure}
\caption{Statistical analysis of ratings for trusted and distrusted reviews defined by historical credibility and helpfulness vote.}
\label{fig:fig15}
\end{figure}

Fig.~\ref{fig:fig15} compares the mean and standard deviation of the ratings for trusted and distrusted reviews defined by historical credibility and the helpfulness vote. It also demonstrates the consistent result that the trusted reviews have a lower mean and higher standard deviation compared to the distrusted reviews in all the definitions for historical credibility. 
Similar to Fig.~\ref{fig:fig14}~(d), the trusted and distrusted reviews classified by the helpfulness vote do not reveal a clear trend in the mean and standard deviation of the ratings.


\subsection{Correlation Analysis}

\begin{figure}[!h]
\begin{subfigure}[a]{0.5\linewidth}
  \centering
  \includegraphics[width=0.97\linewidth]{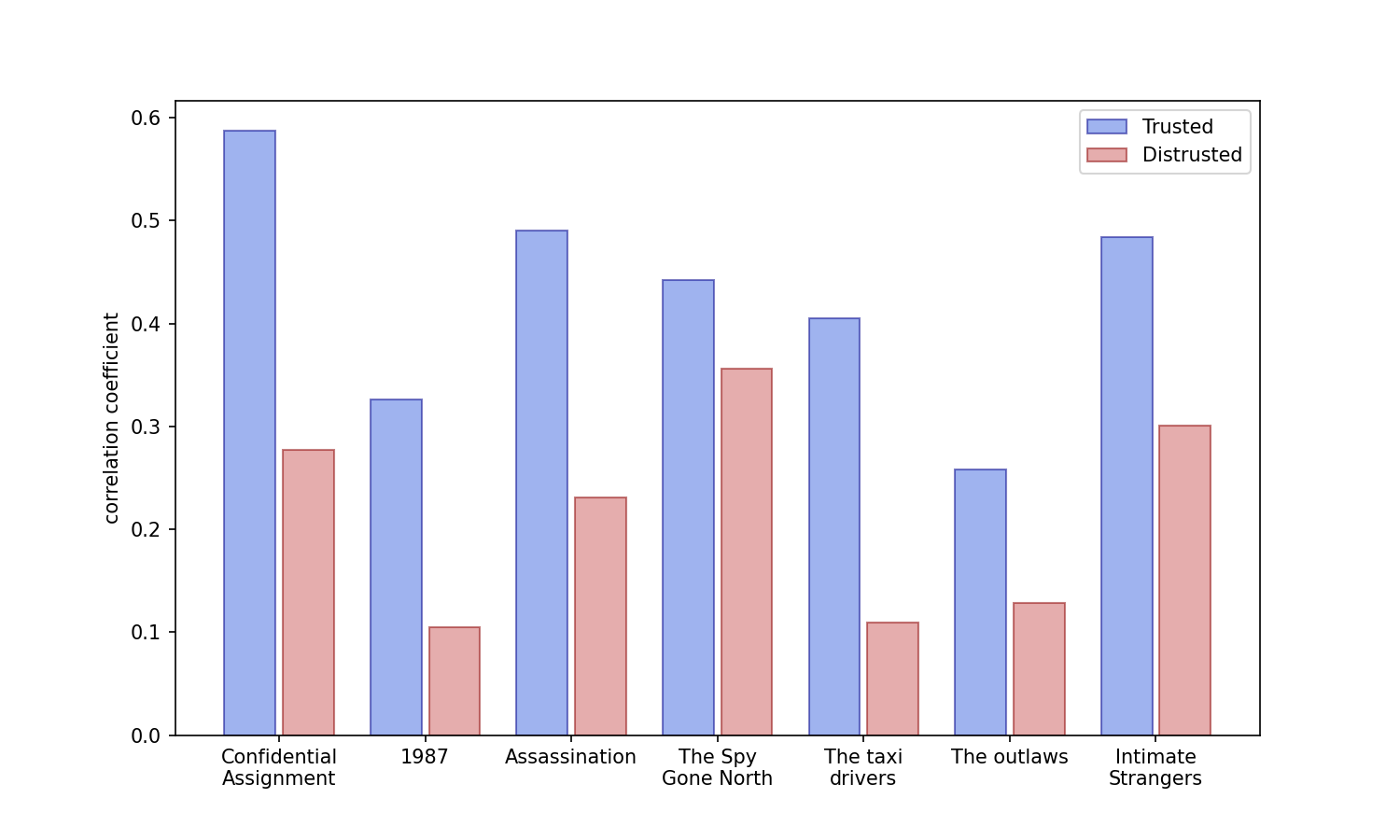}  
  \caption{Definition~1}
  \label{fig:fig5a}
\end{subfigure}
\begin{subfigure}[c]{0.5\linewidth}
  \centering
  \includegraphics[width=0.97\linewidth]{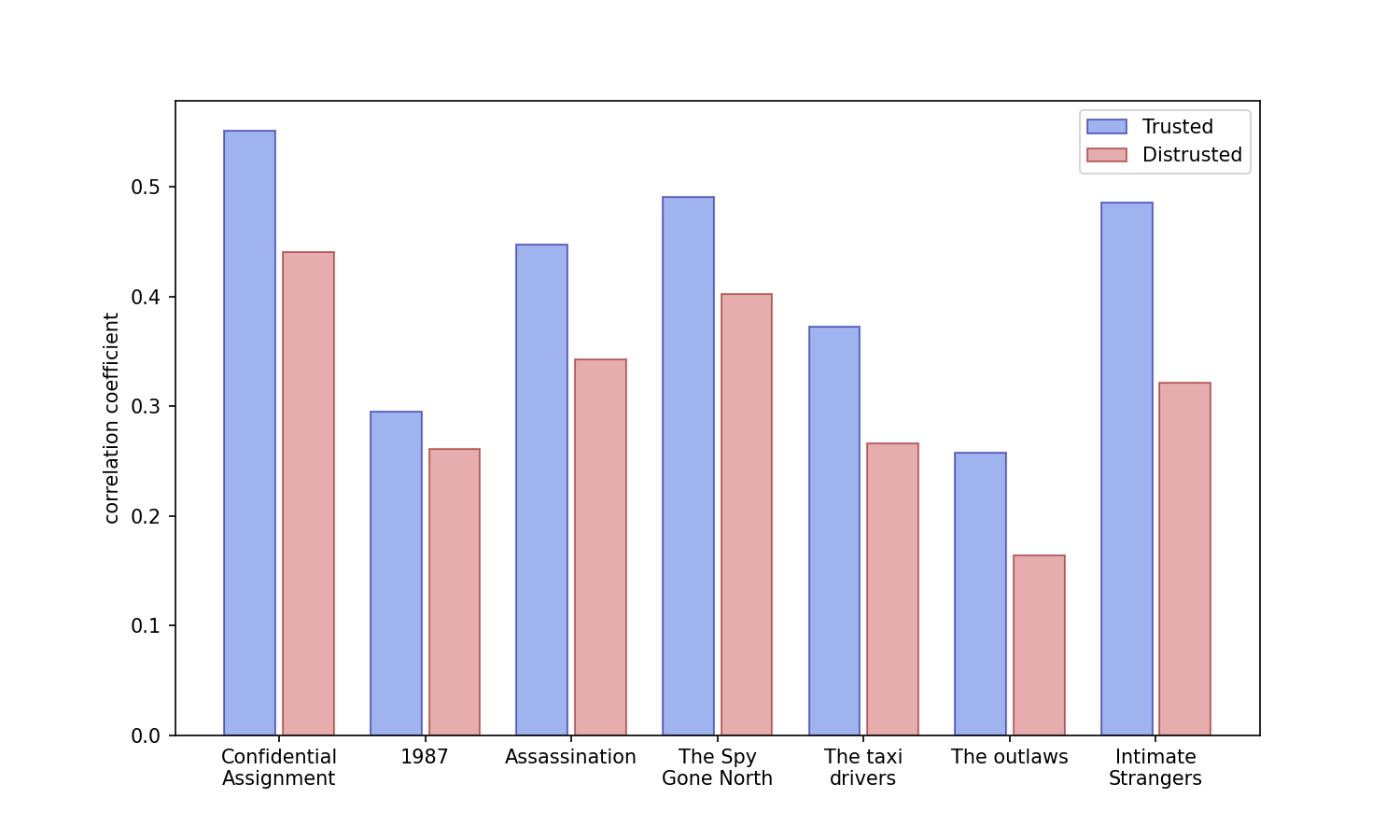}  
  \caption{Definition~2}
  \label{fig:fig5b}
\end{subfigure}
\begin{subfigure}[c]{0.5\linewidth}
  \centering
  \includegraphics[width=0.97\linewidth]{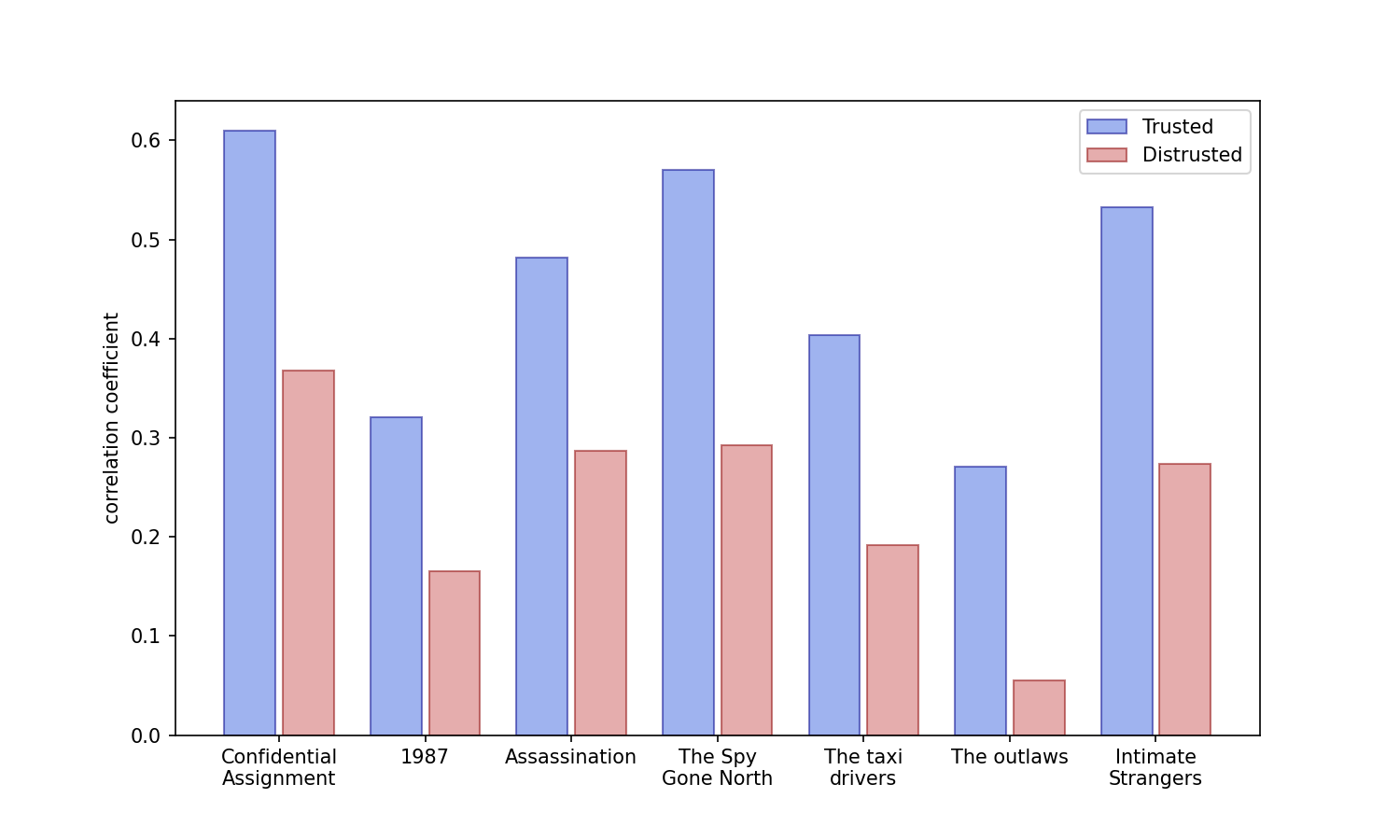}  
  \caption{Definition~3}
  \label{fig:fig5c}
\end{subfigure}
\begin{subfigure}[d]{0.5\linewidth}
  \centering
  \includegraphics[width=0.97\linewidth]{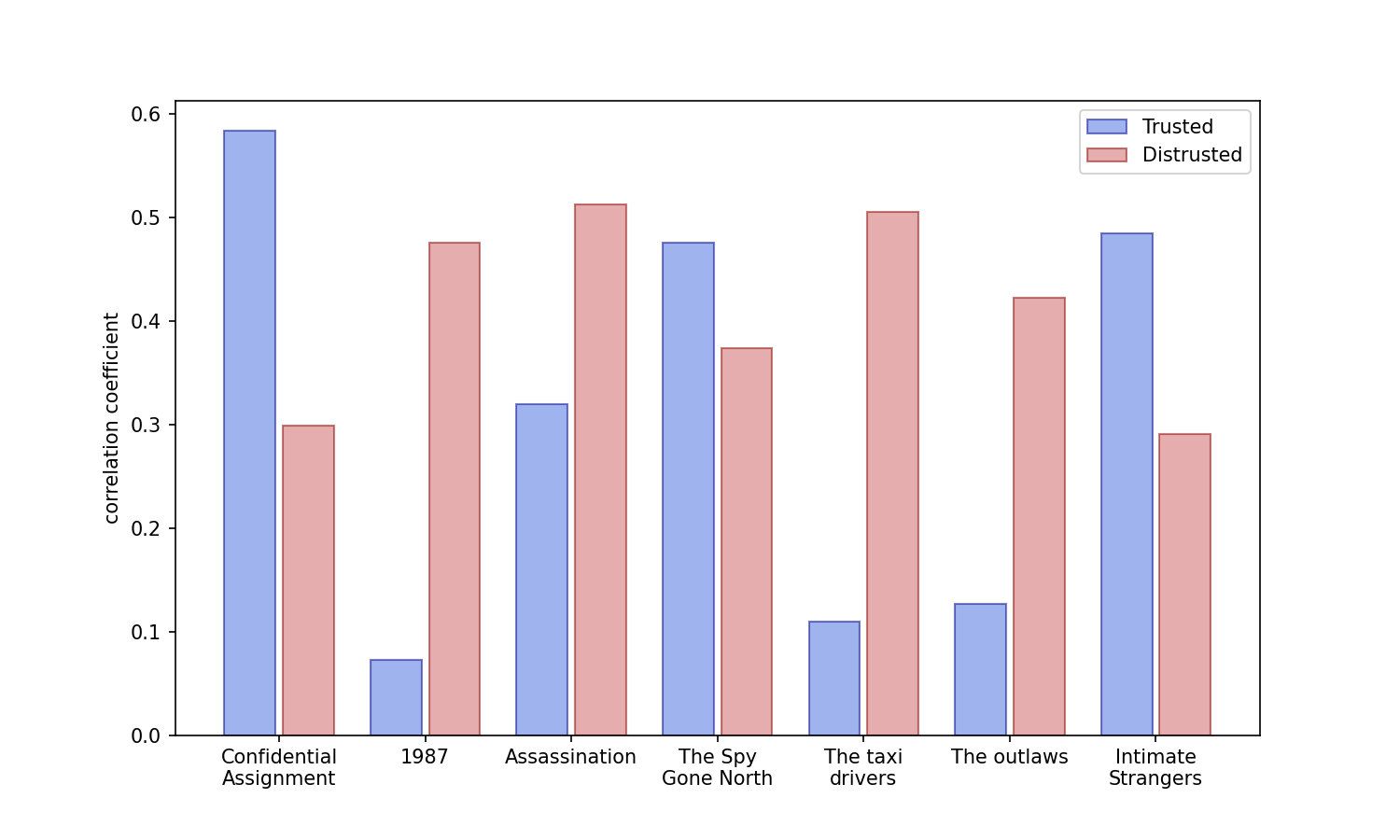}  
  \caption{Helpfulness votes}
  \label{fig:fig5d}
\end{subfigure}
\caption{Correlation analysis for trusted and distrusted reviews defined by historical credibility and helpfulness vote.}
\label{fig:fig16}
\end{figure}

In this section, we compare the difference between trusted and distrusted reviews through correlation analysis between the sentiment scores of the reviews and the ratings. To measure the correlation between two sets of each reviewer, we use  Spearman’s Correlation Coefficient. It allows us to measure the statistical dependence of ranking between two sets and to determine the strength and direction of the association. Fig.~\ref{fig:fig16} compares the results of Spearman’s correlation analysis for the reviews defined by historical credibility and the helpfulness vote. Here, we averaged the correlation coefficient of reviewers in the trusted or distrusted reviewers classified by each criterion. The result indicates that the trusted reviews show a much higher correlation between the reviews and ratings than the distrusted reviews. Specifically, the mean of the correlation coefficients for the trusted reviews defined by Definition~1 is 0.4386, and that for the distrusted reviews is 0.2676; the mean of the correlation coefficients for the trusted review by Definition~2 is 0.4691, and that for the distrusted reviews is 0.2068; the mean of the correlation coefficients for the trusted reviews  defined by Definition~3 is 0.4411, and that for the distrusted reviews is 0.2741. This result indicates a significantly improved consistency between reviews and ratings of the trusted reviewers compared to the distrusted reviewers by the proposed historical credibility, providing evidence to show its effectiveness.

\section{Weakly Supervised Classification based on Historical Credibility}
\label{sec:sec5}
\label{sec:sec4-1}

In this section, we apply the proposed historical credibility to classify a given review as a trusted or distrusted one. For this, we aim to utilize the strength of historical credibility: 1) it can directly determine the review credibility based on the ratings and reviews, instead of accumulating the records such as the helpfulness vote, and 2) it utilizes criteria based on distinguishable statistics obtained from ratings and sentiment scores rather than ambiguous short textual reviews. We apply historical credibility to the weakly supervised classification model. It allows 1) fast annotation by the clear criteria and 2) improvement of the classification accuracy compared to the case using textual reviews.



We define 1) ratings, 2) sentiment scores, and 3) the correlation coefficients between them as the features to learn the classification model, which are essential statistics used for defining historical credibility. 
Then, we annotate the entire movie review as ``trusted'' or ``distrusted'' based on each criterion of historical credibility. Here, we note that the annotation of the entire data set is performed at a very fast speed because we can annotate them using a clear criterion without manual effort. Finally, we build a classification model based on two machine learning models: LSTM and SVM. In the inference phase, we judge the credibility of new movie reviews using the classification model. This implies that we can even judge the credibility of movie reviews that are written by first-time reviewers.

\subsection{Experimental Methods}
\label{sec:sec5-1}



We collect reviews from six movies having ratings and textual reviews. We also collect historical textual reviews and ratings for each reviewer. The textual reviews are described in up to 140 characters. 
Table~\ref{tab:table2} shows the characteristics of the movie reviews collected, which are publicly available at Mendeley Data\footnote{https://data.mendeley.com/datasets/jb5knzh8yv/4}. We collected 38,400 movie reviews, 6,400 for each movie. Here, 30,720 reviews are used for training and 7,680 are used for testing. Here, to avoid the data imbalance problem between the trusted and distrusted reviews, we used the same number of trusted and distrusted reviews.

\begin{table}[]
\centering
\caption{Movie review data sets for the classification model.}
\label{tab:table2}
\begin{tabular}{|c|c|c|c|}
\hline
\textbf{Movies}                                               & \textbf{Release date} & \textbf{\begin{tabular}[c]{@{}c@{}}Training \\ data\end{tabular}} & \textbf{\begin{tabular}[c]{@{}c@{}}Testing\\ data\end{tabular}}                                                               \\ \hline \hline
 \begin{tabular}[c]{@{}c@{}}Intimate Strangers\end{tabular}                                           & 31 Oct. 2018          & 5,120                                                                     & 1,280                       
\\ \hline A Taxi Driver   & 02 Aug. 2017  & 5,120   & 1,280                                                                    
\\ \hline
Assassination       & 22 Jul. 2015          & 5,120              & 1,280                                                                    
\\ \hline
The Spy Gone North      & 08 Aug. 2018          & 5,120       & 1,280                                                                    
\\ \hline
1987    & 27 Dec. 2017          & 5,120  & 1,280                                                                    
\\ \hline
 Confidential Assignment    & 18 Jan. 2017    & 5,120             & 1,280                                                                    \\ \hline
\hline
\multicolumn{2}{|c|}{Total}                                                                                    & 30,720                                                                      & 7,680                                                                    \\ \hline
\end{tabular}
\end{table}


In the experiments, we measure the elapsed learning time to check the efficiency of annotation in the proposed learning model. In addition, we measure the accuracy of the proposed classification model based on the features in historical credibility and that of the textual review-based learning model. 




\subsection{Learning time}
\label{sec:sec5-3}

Table~\ref{tab:tab4} shows the average elapsed learning time of the proposed learning method for the six movies by the machine learning model. Here, to measure the elapsed time, we use 6,400 movie reviews for each movie. We note that annotation of 6,400 movie reviews requires only 0.093 seconds on average, which only occupies 0.55\% and 1.88\% of the total learning time for LSTM and SVM, respectively.

\begin{table}
\fontsize{9}{10}\selectfont
\center{
\caption{Elapsed learning time for movie reviews (seconds).} 
\setlength{\tabcolsep}{20pt}
\begin{tabular}{|c|c|c|}

\hline
Model & \textbf{LSTM} & \textbf{SVM}  \\
\hline
\hline Annotation & \multicolumn{2}{|c|}{0.093}  \\ 
\hline Training & 14.043 & 3.532 \\
\hline Testing & 2.621 & 1.317  \\
\hline
\hline  \textbf{Total} & 16.757 & 4.942 \\

\hline
\end{tabular}
\label{tab:tab4}
}
\end{table}

\subsection{Accuracy}

\begin{figure}[!h] 
  \begin{subfigure}[b]{0.33\linewidth}
    \centering
    \includegraphics[width=0.98\linewidth]{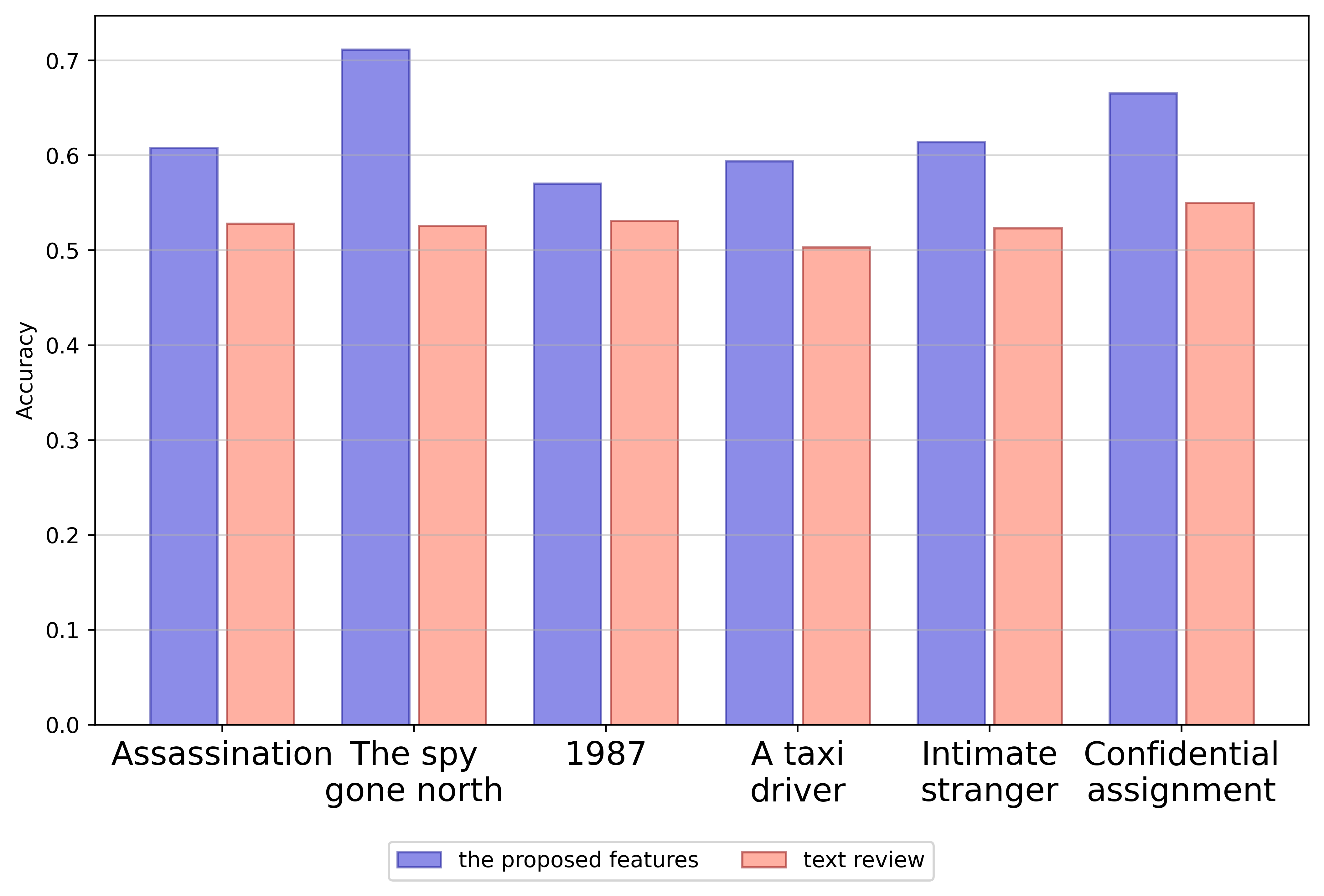} 
    \caption{Definition~1.} 
    \label{fig:fig6-a} 
  \end{subfigure}
  \begin{subfigure}[b]{0.33\linewidth}
    \centering
    \includegraphics[width=0.98\linewidth]{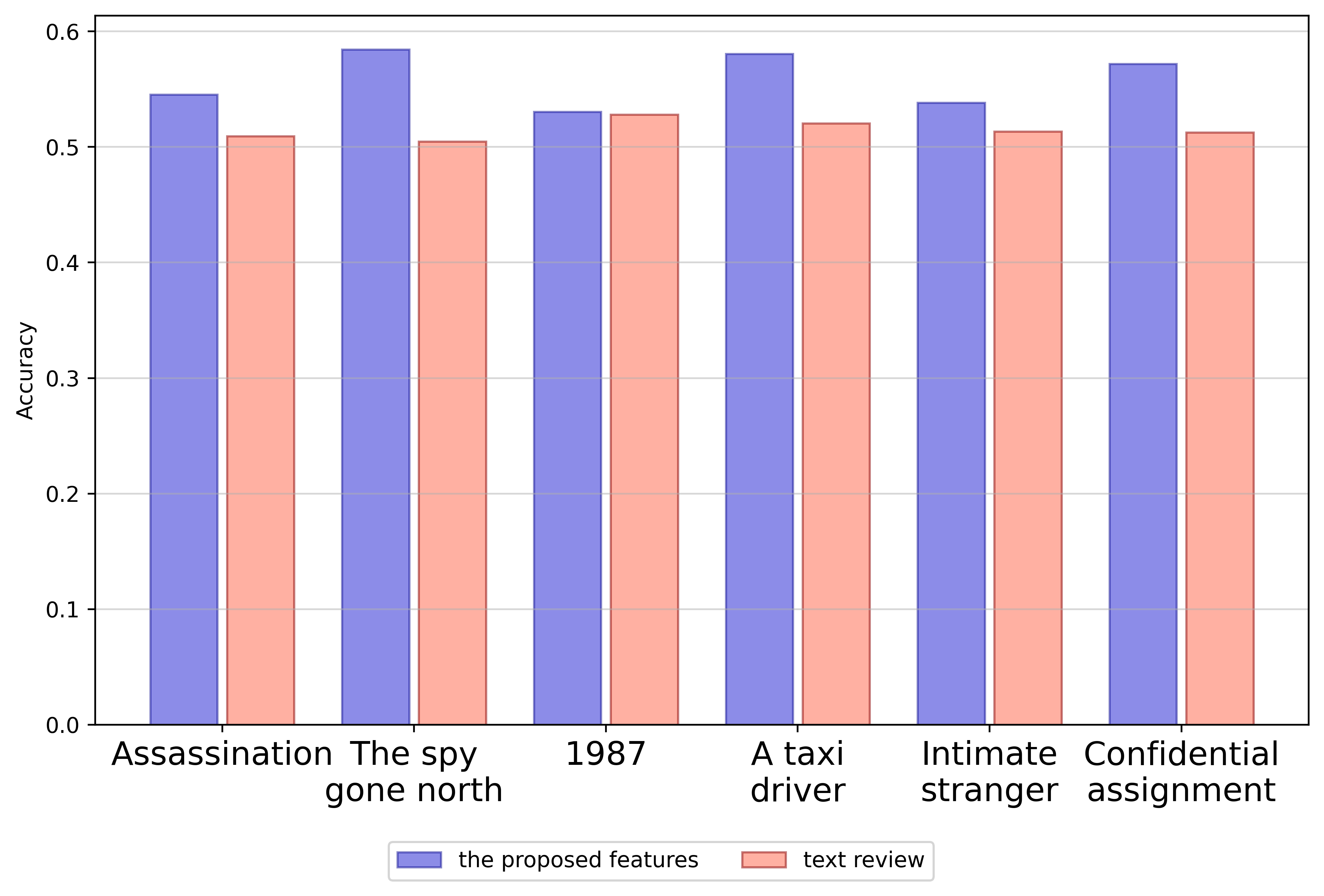} 
    \caption{Definition~2.}
    \label{fig:fig6-b} 
  \end{subfigure} 
  \begin{subfigure}[b]{0.33\linewidth}
    \centering
    \includegraphics[width=0.98\linewidth]{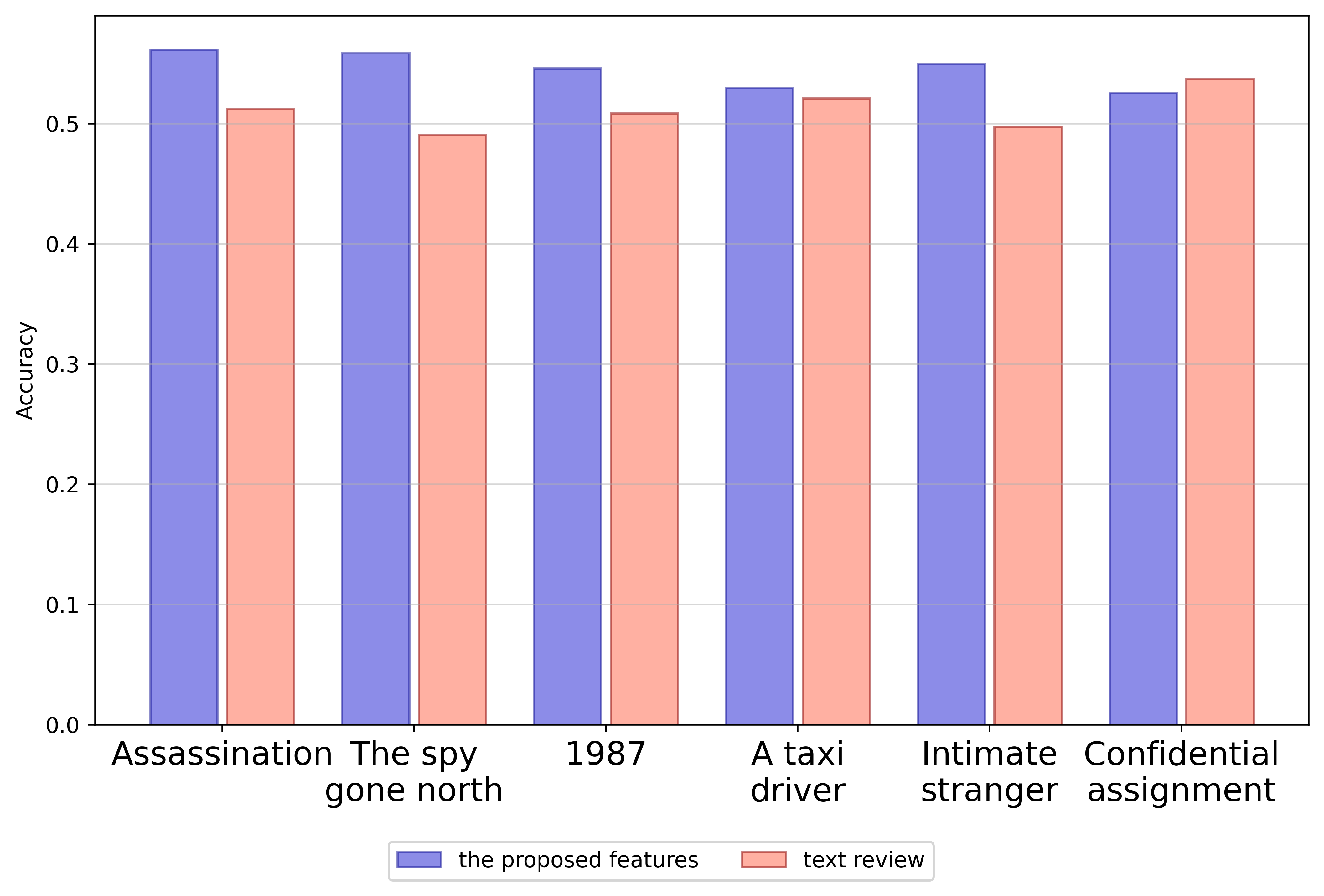} 
    \caption{Definition~3.} 
    \label{fig:fig6-c} 
  \end{subfigure}
  \caption{Accuracy comparison between historical credibility-based LSTM and textual review-based LSTM.}
  \label{fig:fig6} 
 
\end{figure}

Fig.~\ref{fig:fig6} compares the accuracy between the historical credibility-based LSTM and textual review-based LSTM. The results indicate that the proposed classification model clearly outperforms the textual review-based model for all the definitions. Specifically, the former outperforms the latter by up to 11.4\% in the case of Definition~1.

\begin{figure}[!h] 
  \begin{subfigure}[b]{0.33\linewidth}
    \centering
    \includegraphics[width=0.98\linewidth]{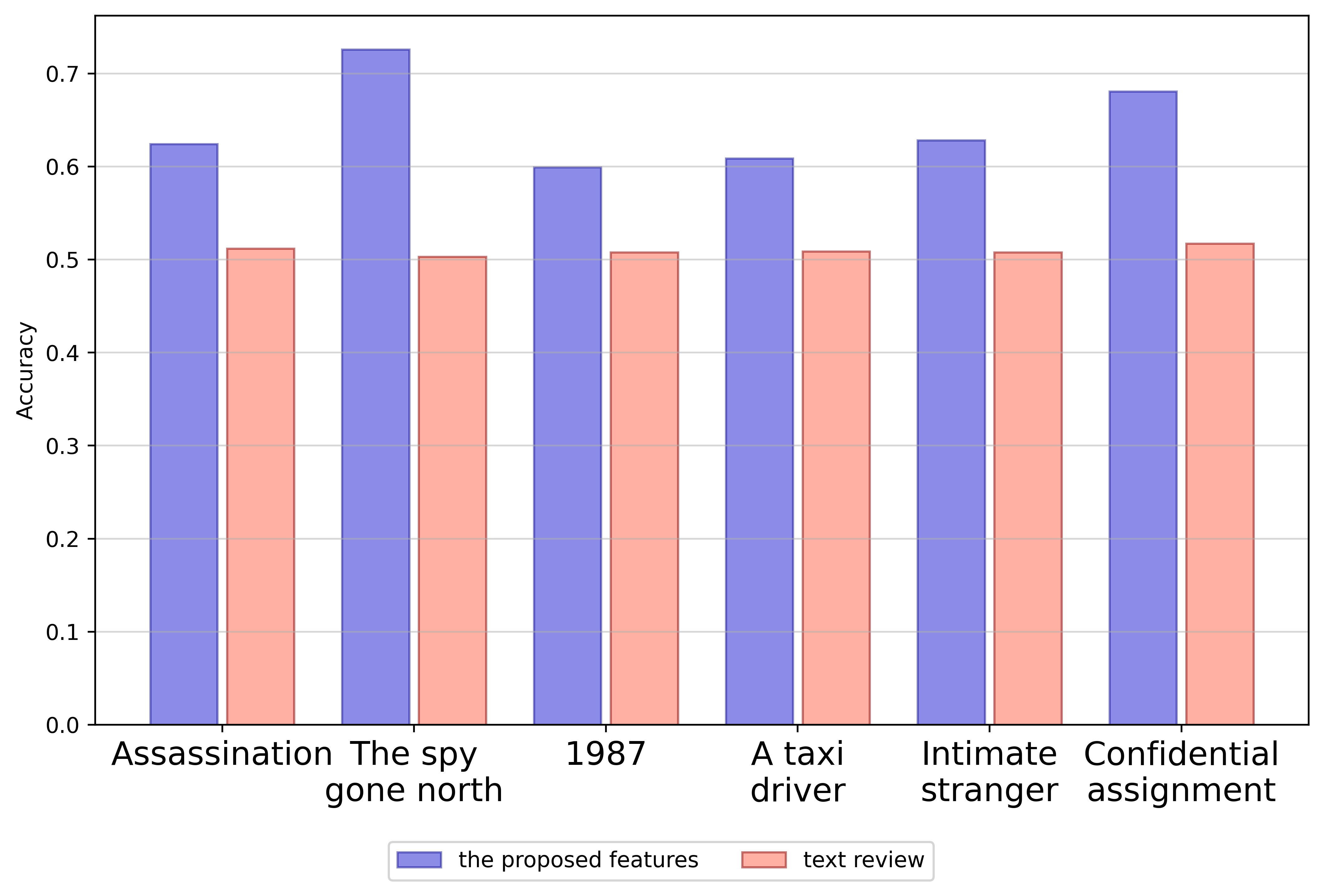} 
    \caption{Definition~1} 
    \label{fig:fig9-a} 
  \end{subfigure}
  \begin{subfigure}[b]{0.33\linewidth}
    \centering
    \includegraphics[width=0.98\linewidth]{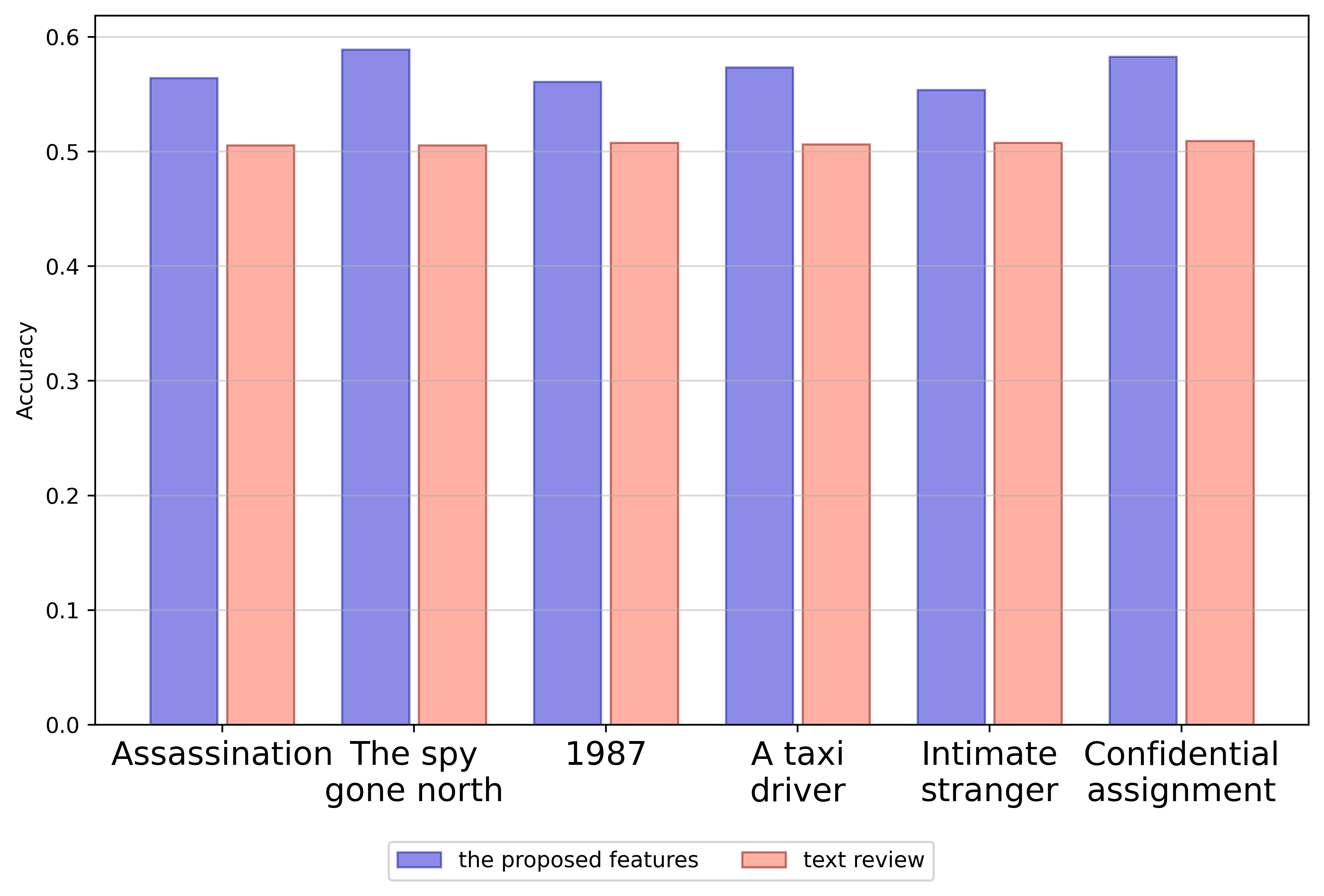} 
    \caption{Definition~2} 
    \label{fig:fig9-b} 
  \end{subfigure} 
  \begin{subfigure}[b]{0.33\linewidth}
    \centering
    \includegraphics[width=0.98\linewidth]{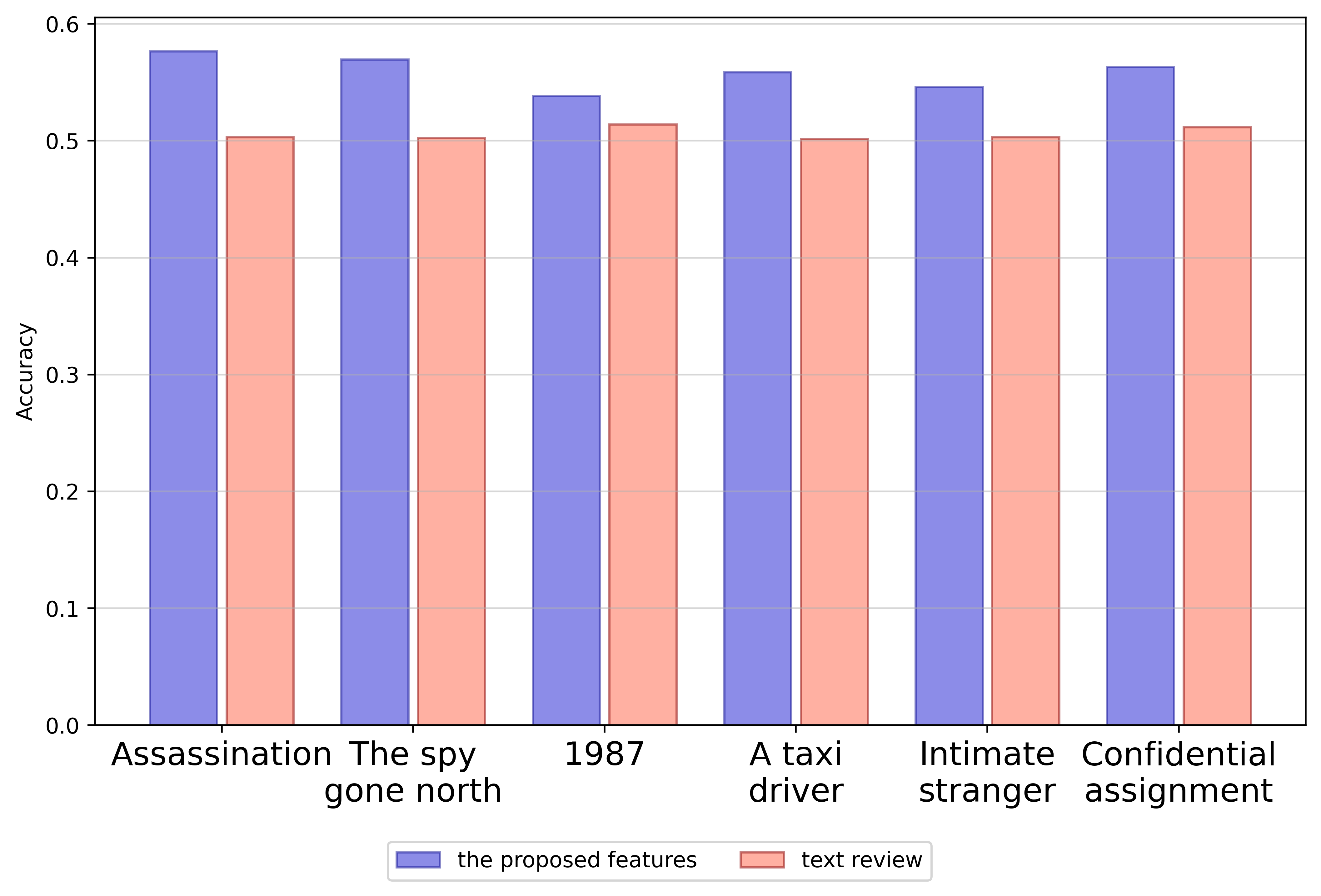} 
    \caption{Definition~3} 
    \label{fig:fig9-c} 
  \end{subfigure}
  \caption{Accuracy comparison between the historical credibility-based SVM and textual review-SVM.}
  \label{fig:fig9} 
 
\end{figure}

Fig.~\ref{fig:fig9} compares the accuracy between the historical credibility-based SVM and textual review-based SVM. Similar to Fig.~\ref{fig:fig6}, the results indicate that the proposed classification model outperforms the textual review-based classification model for all the definitions. Specifically, the former outperforms the latter by up to 13.7\% in the case of Definition~1. 


\begin{figure}[!h] 
  \begin{subfigure}[b]{0.33\linewidth}
    \centering
    \includegraphics[width=0.98\linewidth]{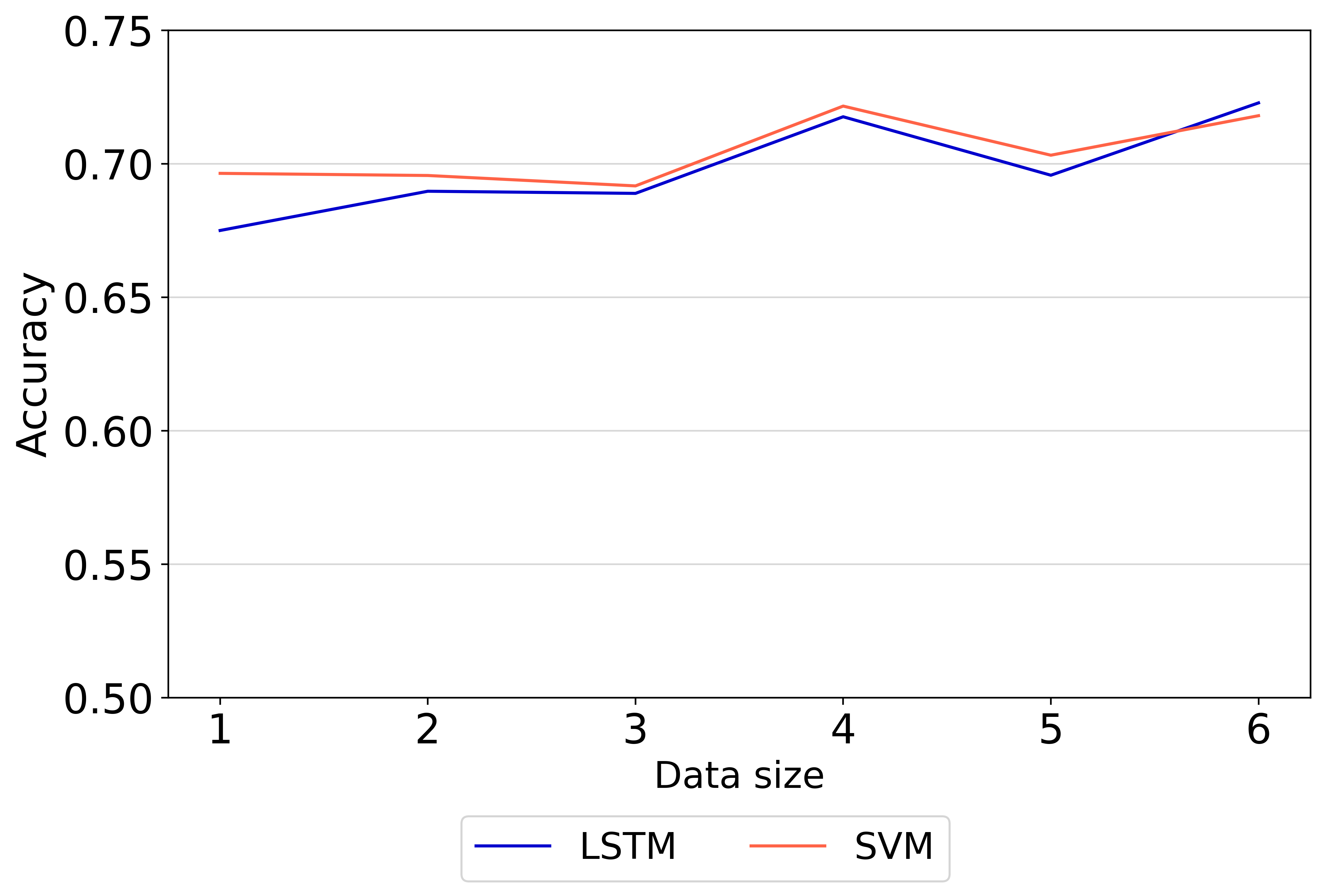} 
    \caption{Definition~1} 
    \label{fig:fig11-a} 
  \end{subfigure}
  \begin{subfigure}[b]{0.33\linewidth}
    \centering
    \includegraphics[width=0.98\linewidth]{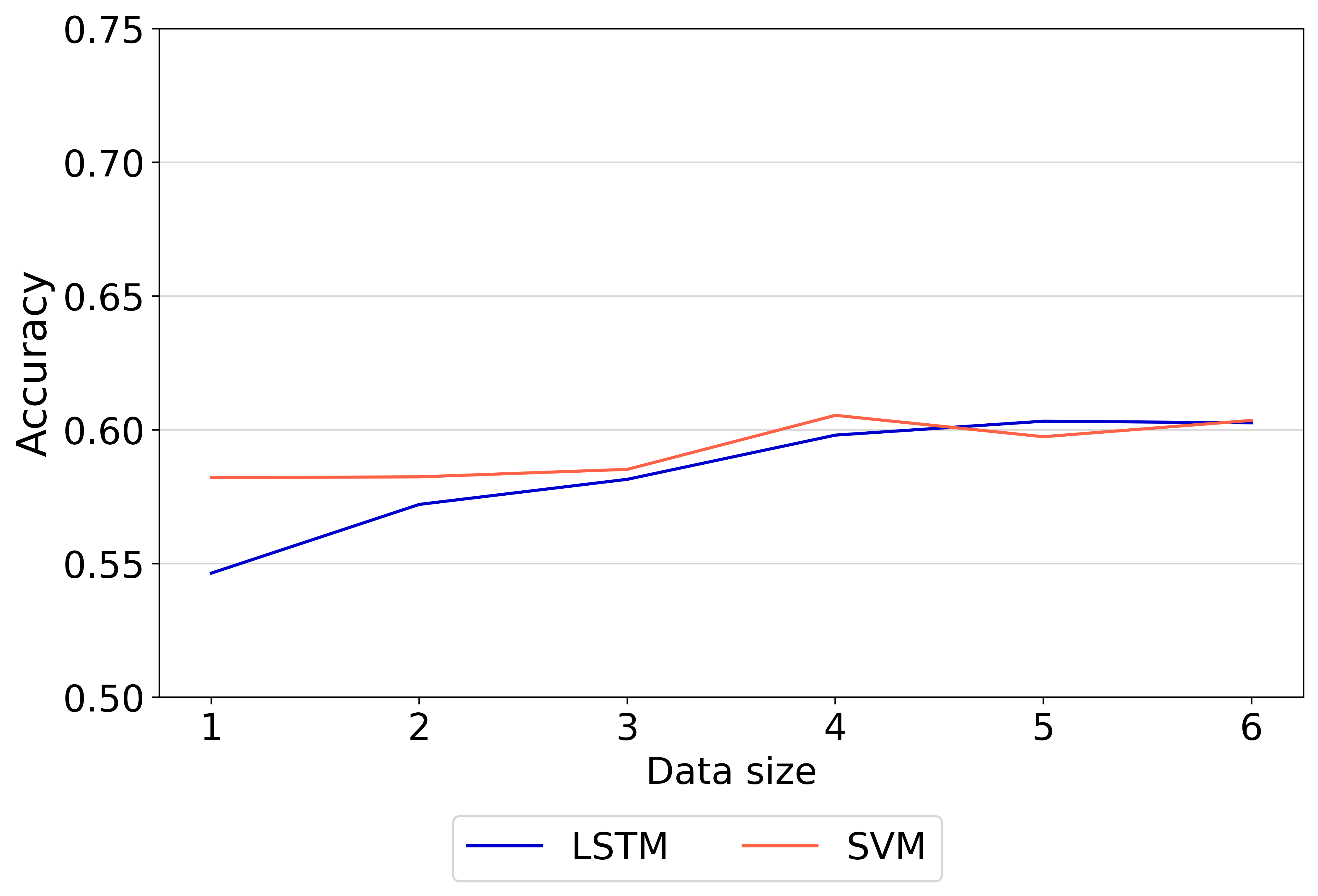} 
    \caption{Definition~2} 
    \label{fig:fig11-b} 
  \end{subfigure} 
  \begin{subfigure}[b]{0.33\linewidth}
    \centering
    \includegraphics[width=0.98\linewidth]{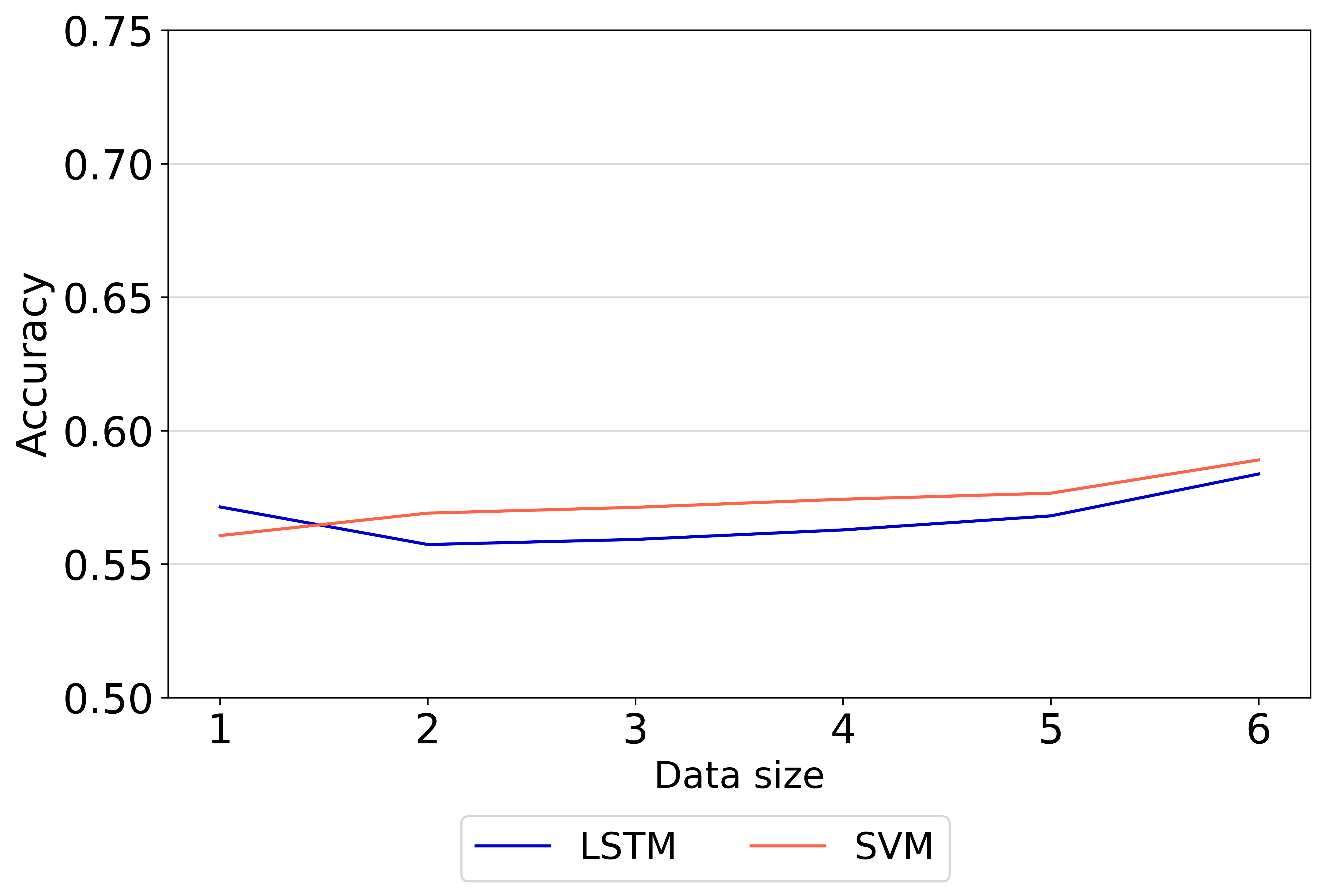} 
    \caption{Definition~3} 
    \label{fig:fig11-c} 
  \end{subfigure}
  \caption{The accuracy of the proposed learning model as data size increases.}
  \label{fig:fig11} 
 
\end{figure}

To check the effectiveness of the proposed classification model, we validate its accuracy as the data set size increases. Fig.~\ref{fig:fig11} shows the accuracy of the proposed classification model by the definition as the data size increases from 1,400 to 11,400. We confirm that the accuracy for all the definitions in both SVM and LSTM clearly becomes higher as the data size increases. Specifically, in Definition~1, the model's accuracy increases by up to 7.08\% in LSTM and by up to  3.1\% in SVM; in Definition~2, the model's accuracy increases by up to 10.3\% in LSTM and by up to 3.67\% in SVM; in Definition~3, the model's accuracy increases by up to 2.16\% in LSTM and by up to 5.05\% in SVM. This result clearly indicates the effectiveness of our classification model because it becomes accurate as the data set is accumulated.

\section{Discussions}
\label{sec:sec6}

\noindent {\bf Accuracy of the classification model: } In this study, we focus on proposing simple, but effective criteria, resulting in that we can apply the criteria to the weakly supervised classification model. With the strengths of weakly supervised learning, we can annotate the movie reviews at a very fast speed and effectively classify the movie reviews using a very simple rule. Naturally, the purpose of our classification model is not to show its high accuracy, but to show the possibility of its improvement for efficiency and accuracy as the data size increases compared to the textual reviews, which are inherently difficult to judge their credibility only considering the textual reviews.

\vspace*{0.2cm}
\noindent {\bf Language-independent method: } In this paper, we use movie reviews collected from the largest Korean movie review website, Naver Movie, because it satisfies two requirements to show the effectiveness of the proposed method: 1) it maintains all the historical movie reviews written by each user, as required for the proposed historical credibility; 2) it maintains helpfulness and unhelpfulness votes for each review, as required for the comparison method. We can apply the proposed method to movie reviews collected from other websites only if they satisfy the two requirements above. We note that the proposed method is language-independent. The accuracy of the classification model might vary according to the languages in movie reviews. In this paper, we have focused on the relative accuracy difference between methods, not on the absolute value of the accuracy depending on the target language. This relative difference should be maintained in other languages.

\vspace*{0.2cm}

\noindent {\bf Machine learning techniques: } In this paper, we aim to show the effectiveness of the proposed classification model, rather than the superiority of a specific machine learning technique. For this reason, we choose two representative machine learning techniques. However, we can easily apply any other machine learning techniques to the proposed framework.

\section{Conclusions}
\label{sec:sec7}


In this study, we have proposed simple and effective criteria for judging the credibility of movie reviews, which we call \emph{historical credibility}. We have validated its effectiveness through extensive analysis. Specifically, we have classified the entire review into the trusted and distrusted reviews by historical credibility and have shown that their characteristics are quite distinguishable in terms of three viewpoints: 1) distribution, 2) statistics, and 3) correlation. The analysis has provided the evidence showing that historical credibility is effective. Specifically, the distribution and statistical analysis have shown that distrusted reviews tend to be more positive and indistinguishable regardless of the movies than trusted reviews. Correlation analysis has shown that the trusted reviews are significantly more consistent in ratings and reviews' sentiment scores than the distrusted reviews. 

We have then applied historical credibility to classify a given movie review as a trusted or distrusted one based on weakly supervised learning. We have shown that the proposed learning method is quite fast because it can annotate 6,400 movie reviews in only 0.093 seconds, which occupies only about 0.55\% $\sim$ 1.88\% of the total learning time. We have shown that the proposed classification model based on historical credibility outperforms the textual review-based model. Specifically, the classification accuracy of the former outperforms that of the latter by up to 11.7\% $\sim$ 13.4\%.
We have also shown that the proposed classification model gradually improves the accuracy when the data size increases.


\end{document}